\documentclass[prd,amsmath,amssymb,nofootinbib,floatfix,superscriptaddress]{revtex4}

\usepackage{slashed} 
\usepackage{physics, xcolor, bbold}
\usepackage{graphicx,bm}
\usepackage[normalem]{ulem} 
\usepackage{amsmath,amssymb}
\usepackage[normalem]{ulem}
\usepackage{hyperref}

\newcommand{\refcite}{\cite}

\usepackage{graphicx,bm,xcolor, bbold,amsmath,amssymb}
\usepackage[normalem]{ulem}
\usepackage{hyperref}

\begin{document}

\title{
  Scaling behaviors at quantum and classical first-order transitions}

 \author{Andrea Pelissetto}
 \altaffiliation{andrea.pelissetto@roma1.infn.it}
\affiliation{Dipartimento di Fisica dell'Universit\`a di Roma Sapienza
        and INFN Sezione di Roma I, I-00185 Roma, Italy}

\author{Ettore Vicari}
 \altaffiliation{ettore.vicari@unipi.it}
\affiliation{Dipartimento di Fisica dell'Universit\`a di Pisa
  and INFN Largo Pontecorvo 3, I-56127 Pisa, Italy}

\begin{abstract}
  We consider quantum and classical first-order transitions, at
  equilibrium and under out-of-equilibrium conditions, mainly focusing
  on quench and adiabatic protocols. For these phenomena, we review
  the finite-size scaling theory appropriate to describe the general
  features of the large-scale (and long-time for dynamic phenomena)
  behavior of finite-size systems.
\end{abstract}

\maketitle

\section{Introduction}
\label{intro}

Classical and quantum phase transitions (PTs) are phenomena of great
interest in modern physics, both theoretically and
experimentally. Classical PTs are generally driven by thermal
fluctuations~\cite{Landau-book,Fisher-65,WK-74,Fisher-74,Ma-book,
  Wilson-83,ZinnJustin-book,Fisher-16}, while their quantum
counterparts arise from quantum fluctuations in the limit of zero
temperature~\cite{SGCS-97,Sachdev-book}.  Transitions separate
qualitatively different phases of matter, each one of them being
characterized by distinctive properties.  In many cases PTs are
associated with the spontaneous breaking of a global symmetry and the
emergence of corresponding ordering phenomena. Some transitions in
systems with gauge symmetries are notable exceptions. In this case,
the transitions are not characterized by a local order parameter
associated with a global symmetry breaking, but are driven by
topological excitations
\cite{Wegner-71,Wen-book,Sachdev-16,Sachdev-19}.

At PTs, in the infinite-volume thermodynamic limit, the free energy,
or, in the case of zero-temperature quantum phase transitions (QPTs),
the low-energy properties become nonanalytic functions of the system
parameters, such as the temperature, the coupling constants,
etc.~\cite{Ruelle-63,Fisher-64,Fisher-65,WK-74, Fisher-74, Ma-book,%
  BGZ-76, Wegner-76, Wilson-83, Fisher-86, Binder-87, Fisher-98,
  PV-02, SGCS-97, Sachdev-book,ZinnJustin-book,Fisher-16,RV-21}.
Depending on the nature of the nonanalyticity at the transition point,
PTs are generally distinguished as first-order or continuous PTs.
They are continuous when the bulk properties change continuously at
the transition point, and the correlation functions develop a
divergent length scale. They are of first order when the thermodynamic
or ground-state properties in the infinite-volume (thermodynamic)
limit are discontinuous across the transition point. See, for example,
Ref.~\refcite{Binder-87} for an introductory review of the main
features of classical first-order PTs at finite temperature.

As already remarked in Ref.~\refcite{Binder-87}, the behaviors
emerging at first-order PTs are more complex than those observed at
continuous transitions. Critical phenomena at continuous PTs are
essentially related to the presence of critical correlations, which
decay as a power of the distance at the critical point, and of a
diverging length scale $\xi$.  When approaching the critical point,
the long-distance behavior, i.e., on distances of the order of the
scale $\xi$, shows universal features that only depend on few global
properties of the microscopic short-distance interactions. The
behavior at first-order transitions is instead more complex. Indeed,
we typically observe the coexistence of different phases, with one or
more ordered phases, i.e., of phases in which the system variables are
correlated on all length scales, apart from some short-range local
fluctuations.  The presence of coexisting phases gives rise to
peculiar competing phenomena in the phase-coexistence region, such as
metastability, nucleation, droplet formation, coarsening,
etc.~\cite{Binder-87,Bray-94} Moreover, the observed bulk behavior
crucially depends on the nature of the boundary conditions (BC), at
variance with what happens at continuous transitions, where BC only
affect some finite-size properties of the system, but are irrelevant
in the thermodynamic limit.

In general, the singular behavior at PTs only occurs in the
infinite-volume limit.  If the size $L$ of the system is finite, all
properties are analytic as a function of the external parameter
driving the transition. However, close to the transition point and for
large system sizes, the finite-size behavior of thermodynamic
quantities and long-distance properties shows some general scaling
features.  At continuous transitions statistical systems develop
finite-size scaling (FSS) behaviors when $L$ is large and the length
scale $\xi$ of the critical modes is comparable with the size $L$ of
the system. More precisely, this occurs in the FSS limit, defined as
the limit $L,\xi\to\infty$ keeping the ratio $L/\xi$ fixed. This
regime presents universal features, shared by all systems whose
transition belongs to the same universality
class~\cite{FB-72,Barber-83,Privman-90,Cardy-editor,PV-02,CPV-14,RV-21}.
Although originally formulated in the classical
framework~\cite{FB-72}, FSS also holds at continuous quantum
transitions (CQTs).\cite{SGCS-97,CPV-14,RV-21} Peculiar finite-size
behaviors emerge at classical and quantum first-order transitions as
well~\cite{NN-75,FB-82,PF-83,FP-85,CLB-86,Binder-87,
  BK-90,LK-91,BK-92,VRSB-93,CNPV-14,PRV-18,RV-21}. At variance with
continuous transitions, the qualitative behavior varies with the BC,
since boundaries drastically affect the bulk behavior of the system
(for instance, they may induce domain walls).  The sensitivity to the
BC represents one of the main qualitative differences between
continuous and first-order PTs.  Understanding finite-size effects in
systems close to PTs is essential to correctly interpret experimental
or numerical data, when phase transitions are investigated in
relatively small systems~\cite{FB-72,Barber-83, Binder-87,
  Cardy-editor, Privman-90, PV-02, GKMD-08, CPV-14, CNPV-14, PRV-18,
  RV-21}, or in particle systems trapped by external forces entailing
an additional length scale, as in experiments with cold atoms confined
by an external potential~\cite{BDZ-08,CV-09,CV-10}.

In this review we analyze several aspects characterizing quantum and
classical first-order transitions (FOQT and FOCT, respectively), in
equilibrium and out-of-equilibrium conditions. Our presentation is
meant to give a unified picture of quantum and classical first-order
transitions, emphasizing the deep connections between them, which can
be understood using the so-called quantum-to-classical
mapping~\cite{Sachdev-book,CPV-14,RV-21}, but also stressing their
crucial differences, which are due to the different nature of the
fluctuations driving the transitions.  We discuss the finite-size
behavior at FOQTs and FOCTs, focusing on the interplay between the
thermal and/or quantum fluctuations driving the PT and the size of the
system, and on the role of the BC.  Somewhat unconventionally, we
begin our presentation by reviewing results for FOQTs, as in this case
the FSS behavior emerges more clearly.  Then, we extend the discussion
to FOCTs. In particular, we use the general quantum-to-classical
mapping of $d$-dimensional quantum systems onto $D$-dimensional
($D=d+1$) classical systems in an anisotropic geometry, to relate the
quantum and classical results.

We stress that a good theoretical understanding of the behavior at
first-order transitions is phenomenologically relevant, as they are
ubiquitous. For example, the condensation of water, the melting of
ice, etc.  are crucially important FOCTs.  FOQTs occur in quantum Hall
systems~\cite{PPBWJ-99}, itinerant ferromagnets~\cite{VBKN-99,
  BKV-99}, heavy fermion metals~\cite{UPH-04, Pfleiderer-05, KRLF-09},
SU($N$) magnets~\cite{DK-16, DK-20}, quantum spin
systems~\cite{LMSS-12,CNPV-14, LZW-19}, etc.

The review is organized as follows.~\footnote{Here we provide the list
of acronyms used in this review: PT (phase transition), CQT
(continuous quantum transition), FOQT (first-order quantum
transition), FOCT (first-order classical transition), RG
(renormalization group), FSS (finite-size scaling), BC (boundary
conditions), PBC (periodic BC), ABC (antiperiodic BC), OBC (open BC),
FBC (fixed BC), OFBC (opposite FBC), EFBC (equal FBC).}

$\bullet$ Sec.~\ref{foqt} discusses the equilibrium FSS behavior at
the FOQTs occurring in the paradigmatic quantum Ising and Potts models
on cubic-like lattices. We point out that intensive global quantities
show a universal FSS behavior that can be parametrized in terms of the
parameter driving the FOQT, the volume of the system, and the energy
gap (energy difference of the lowest states of the spectrum). Since
the energy gap has a size dependence that is very sensitive to the BC
(it can decrease exponentially or as a power of the size), the
observed FSS behavior varies significantly with the BC, a specific
feature of FOQTs.

$\bullet$ In Sec.~\ref{foqtdynamics} we show how the equilibrium FSS
results can be extended to out-of-equilibrium phenomena at FOQTs. We
define the FSS variables appropriate to describe the system behavior
when the Hamiltonian parameters are suddenly changed (quench
protocols), or when they are slowly (quasi-adiabatically) changed
across the transition, as in Kibble-Zurek
protocols~\cite{Kibble-80,Zurek-96}.  The time scale of these
out-of-equilibrium processes is also very sensitive to BC. In
particular, if the BC are such that the spectrum is characterized by a
gap that is exponentially small in the size of the system, the dynamic
FSS functions can be obtained from a simpler quantum-mechanical model
that only encodes the evolution of the lowest-energy states.

$\bullet$ In Sec.~\ref{qtoc} we discuss the quantum-to-classical
mapping, which allows one to relate phenomena in $d$-dimensional
quantum systems to $D$-dimensional $(D=d+1$) classical statistical
systems at equilibrium.

$\bullet$ In Sec.~\ref{sec5} we outline the main features of the
equilibrium FSS behavior at FOCTs, pointing out the analogies between
the behavior of classical and quantum systems.  Again the sensitivity
of the large-scale behavior to the nature of the BC emerges as a
characterizing qualitative feature, distinguishing FOCTs from
classical continuous transitions.

$\bullet$ In Sec.~\ref{sec6} we discuss some selected topics on the
dynamics of classical systems close to FOCTs.  We first discuss the
equilibrium dynamics, focusing on the purely relaxational
dynamics~\cite{HH-77,Ma-book}. Then, we consider two different
out-of-equilibrium dynamics, which are the classical analogue of those
discussed in the quantum setting in Sec.~\ref{foqtdynamics}. In
particular, we present some recent results for the dynamics of the
droplet formation in Potts systems using the Kibble-Zurek dynamics. We
do not discuss coarsening phenomena, as they are already outlined in
Ref.~\refcite{Bray-94}.

\section{First-order quantum transitions}
\label{foqt}

Quantum PTs are striking signatures of many-body collective quantum
behaviors~\cite{SGCS-97, Sachdev-book}.  They are continuous when the
ground state of the system changes continuously at the transition
point and correlation functions develop a divergent length scale. They
are, instead, of first order when the ground-state properties are
discontinuous across the transition point.  FOQTs occur when the
lowest energy states cross in the infinite-volume limit. Since the
ground state is different at the two sides of the transition, the
physical properties change discontinuously.  Note that, in the absence
of conservation laws, a level crossing can only occur in the
infinite-volume limit. In a finite system, the degeneracy is lifted
and the level crossing is avoided.

At FOQTs, the low-energy properties are particularly sensitive to the
BC, giving rise to a variety of behaviors, which are even more diverse
than those occurring at CQTs.\cite{SGCS-97, Sachdev-book, CPV-14,
  RV-21} Indeed, depending on the BC---for example, whether they are
neutral or favor one of the phases---one can obtain qualitatively
different static and dynamic
properties~\cite{CNPV-14,CNPV-15,CPV-15,PRV-18,YCDS-18,RV-18,PRV-20}.

To understand the origin of the variety of behaviors observed at
FOQTs, it is useful to discuss the size behavior of the gap $\Delta$,
which is the difference of the energies of the lowest Hamiltonian
eigenstates, since it provides the relevant energy scale at the
transition.  At CQTs, $\Delta$ always scales as $L^{-z}$ where $z$ is
the universal dynamic exponent, independently of the BC~\cite{CPV-14},
which only determine the amplitude of the asymptotic power law. On the
other hand, at FOQTs the size dependence of $\Delta$ varies
significantly with the BC~\cite{CNPV-14, CPV-15, CPV-15-iswb}.  For
some BC the gap decays exponentially with increasing $L$, as
$\Delta\sim e^{-cL^d}$.  Instead, $\Delta$ scales as $L^{-\kappa}$
(where $\kappa$ is a positive power that is usually an integer
number), when BC favor the presence of domain walls, that separate the
space into distinct regions associated with different phases.  The
sensitivity of the large-size behavior of $\Delta$ and, in general, of
all low-energy properties of the system to the BC may be considered as
the main difference between the behaviors of finite-size systems at
CQTs and FOQTs.

In the following we discuss the main features of the FSS behavior at
FOQTs using the quantum Ising and Potts models as paradigmatic models.
We first discuss FSS for Ising systems with neutral BC, where one can
predict the large-scale behavior using an effective model in which
only the relevant low-energy states are considered.  Then, we consider
BC that give rise to domain walls, obtaining a substantially different
picture, and BC that favor one of the two phases. Analogous issues are
discussed in quantum Potts models.

\subsection{Quantum Ising model}
\label{quisi}

\subsubsection{Definition of the model}

To make our discussion concrete, we consider the paradigmatic quantum
$d$-dimensional Ising model on a $d$-dimensional cubic-like lattice of
linear size $L$.  The Hamiltonian is
\begin{equation}
  \hat H = - J \sum_{\langle {\bm x},{\bm y}\rangle}
  \hat \sigma^{(1)}_{\bm x} \hat \sigma^{(1)}_{\bm y} - g \sum_{\bm x}
  \hat \sigma^{(3)}_{\bm x} - h \sum_{\bm x} \hat \sigma^{(1)}_{\bm x}\,,
  \label{hisdef}
\end{equation}
where $\hat \sigma^{(k)}$ are the usual spin-1/2 Pauli matrices
($k=1,2,3$), the first sum is over all bonds connecting
nearest-neighbour sites $\langle {\bm x},{\bm y}\rangle$, and the
other two sums are over all sites ${\bm x}$.  The parameters $g$ and
$h$ represent the external homogeneous transverse and longitudinal
fields, respectively.  Without loss of generality, we assume $J=1$,
$g>0$, and a lattice spacing $a=1$. For $h = 0$, the infinite-volume
Ising model is symmetric under the ${\mathbb Z}_2$ transformation
$\sigma^{(i)}_{\bm x} \to U \sigma^{(i)}_{\bm x} U^\dagger$ with $U =
\prod_{\bm x} \sigma^{(3)}_{\bm x}$, that maps $\sigma^{(1)}_{\bm x}
\to -\sigma^{(1)}_{\bm x}$ and $\sigma^{(3)}_{\bm x} \to
\sigma^{(3)}_{\bm x}$.

For $h=0$, the model undergoes a CQT at a critical point $g=g_c$ ($g_c
= 1$ in one dimension), separating a disordered phase ($g>g_c$) from
an ordered one ($g<g_c$).  The order parameter is the longitudinal
magnetization $m_{\bm x}\equiv \langle \Psi_0(g,h) | \hat \sigma_{\bm
  x}^{(1)}| \Psi_0(g,h)\rangle$, where $|\Psi_0(g,h)\rangle$ is the
ground state for the given $g$ and $h$.  The parameters $r\equiv
g-g_c$ and $h$ represent the (even and odd, respectively) relevant
perturbations of the critical behavior at the CQT. For $g<g_c$, the
longitudinal field $h$ drives FOQTs, where the (average) longitudinal
magnetization $M$ becomes discontinuous in the infinite-volume limit:
\begin{equation}
  \lim_{h \to 0^\pm} \lim_{L\to\infty} M = \pm m_0\,,
  \quad M \equiv L^{-d} \sum_{\bm x} m_{\bm x}\,,
  \quad m_{\bm
  x}\equiv \langle \Psi_0(g,h) | \hat \sigma_{\bm x}^{(1)}|
\Psi_0(g,h)\rangle\, .
  \label{disco}
\end{equation}
The FOQT separates two different phases characterized by opposite
values of the magnetization $m_0$.  The magnetization for $d=1$ is
known~\cite{Pfeuty-70}: $m_0 = (1 - g^2)^{1/8}$.

Several BC can be considered: (i) Periodic BC (PBC), for which $\hat
\sigma^{(k)}_{\bm x} = \hat \sigma^{(k)}_{{\bm x}+L\vec{\mu}}$
($\vec{\mu}$ indicates a generic lattice direction); (ii) Antiperiodic
BC (ABC), for which $\hat \sigma^{(k)}_{\bm x} = - \hat
\sigma^{(k)}_{{\bm x}+L\vec{\mu}}$; (iii) Open BC (OBC); (iv) Fixed BC
(FBC), where the states of the spins on the lattice boundaries are
fixed, typically choosing one of the eigenstates of the longitudinal
spin operator $\hat \sigma^{(1)}$; (v) OBC with boundary fields.  For
BC of type (iv) and (v), we should additionally specify the behavior
at the boundary.  In one dimension, one can consider equal FBC
(EFBC)---in this case the spins on the two opposite boundaries are
eigenstates of $\hat \sigma^{(1)}$ with the same eigenvalue---or
opposite FBC (OFBC)---the boundary spins are eigenstates of $\hat
\sigma^{(1)}$ with opposite eigenvalues. Analogously, we may consider
systems with equal or opposite transverse boundary fields.  Note that
the finite system is translation invariant if PBC and ABC are used.
On the other hand, in systems with OBC, FBC or with boundary fields,
translation invariance is lost.

As we shall see, the FSS behavior depends on the behavior of the BC
with respect to the ${\mathbb Z}_2$ symmetry. We distinguish neutral
BC that do not break the ${\mathbb Z}_2$ symmetry, and therefore do
not favor any of the two phases and nonneutral BC. In particular, PBC,
ABC, and OBC are neutral BC along the FOQT line of Ising systems,
while generic FBC are not. In one-dimension OFBC (or, systems with
opposite longitudinal boundary fields) are not ${\mathbb
  Z}_2$-invariant. However, a finite-size system defined on a lattice
with $1 \le x \le L$ is still symmetric under combined ${\mathbb Z}_2$
and space-reflection transformations: under a space reflection,
$\sigma_{x}^{(i)} \to \sigma_{L-x+1}^{(i)}$.  Thus, also OFBC can be
considered as neutral BC.

\subsubsection{The finite-size gap at the quantum transitions of Ising systems}
\label{gapqutr}

Both at CQTs and FOQTs, the relevant energy scale is the gap $\Delta$.
At the CQT, i.e., for $g=g_c$ and $h=0$, the gap $\Delta(L)$ for a
system of size $L$ scales as $\Delta(L)\sim L^{-z}$, with $z=1$ for
any spatial dimension $d$. In particular, the dynamic exponent $z$ is
independent of the BC.  The amplitudes instead depend on the BC.  For
example, in one dimension, the gap $\Delta(L)$ at the critical point
behaves as~\cite{CPV-14, CPV-15, BG-85}
\begin{equation}
  \Delta_{\rm PBC} = {\pi \over 2L} + O(L^{-2})\,, \;\;\; 
  \Delta_{\rm OBC} = {\pi \over L} + O(L^{-2})\,, \;\;\;
  \Delta_{\rm ABC} = {3\pi \over 2L} + O(L^{-2})\,,
  \label{gapiscqt}
\end{equation}
for PBC, OBC, and ABC, respectively.  The behavior changes along the
FOQT line $g<g_c$ and $h=0$.  Here, BC play a crucial role in
determining the FOQT behavior, as they control the bulk behavior of
the system.  We will distinguish three cases: (i) neutral BC with a
magnetized ground state; (ii) neutral BC with domain walls; (iii)
nonneutral BC.

We consider first neutral BC, such that the magnetized states
$|+\rangle$ and $|-\rangle$ are the only relevant low-energy states
for large values of $L$.  This is the case of PBC and OBC.  In this
case, for $L\to \infty$ and $h=0$ the magnetized states $|+\rangle$
and $|-\rangle$ cross at the transition and satisfy
\begin{equation}
  \langle + \! |\hat \sigma_{\bm x}^{(1)} | \! + \rangle = m_0\,,
  \qquad  \langle - | \hat \sigma_{\bm
    x}^{(1)} | - \rangle = -m_0\,,
  \label{sigmasingexp}
\end{equation}
independently of ${\bm x}$.  The difference of the energies of the
ground state and of the higher excited states remains finite in
the infinite-volume limit.  The degeneracy of the magnetized states is
lifted by the longitudinal field $h$.  Therefore, $h = 0$ is a FOQT
point, where the longitudinal magnetization $M$ becomes discontinuous,
see Eq.~(\ref{disco}).  In a finite system of size $L$, the two
lowest-energy states are superpositions of the two magnetized states.
Due to tunneling effects, the gap $\Delta(L,h)$ for $h=0$ is
finite---the level crossing is avoided---but vanishes exponentially as
$L$ increases~\cite{PF-83, CNPV-14},
\begin{equation}
  \Delta(L,h=0) \equiv E_1(L,h=0)-E_0(L,h=0) \sim e^{-c L^d}\,,
  \label{del}
\end{equation}
where $c$ is a positive constant that depends on $g$ (multiplicative
powers of $L$ may also be present).  In particular, for a quantum
Ising chain with $g<1$ and $h=0$, the gap at the transition point
$h=0$ scales as ~\cite{Pfeuty-70, CJ-87}
  \begin{eqnarray}
    \Delta_{\rm OBC}= 2 \, (1-g^2) \; g^L \, [1+ O(g^{2L})]\,,\qquad
    \Delta_{\rm PBC} \approx 2 \,
    \sqrt{1-g^2\over \pi L} \; g^L\,,
        \label{deltaopbc}
  \end{eqnarray}
for OBC and PBC, respectively.  The differences $\Delta_n(L,h) \equiv
E_n(L,h)-E_0(L,h)$ for the higher excited states $n>1$ approach finite
values for $L\to \infty$.

A second class of BC are neutral BC that induce the presence of domain
walls. We consider first the one-dimensional case. This type of
behavior can be observed for OFBC, or for systems with sufficiently
strong oppositely-directed boundary fields. In this case the
magnetized states are no longer the relevant low-energy
states. Instead, the lowest-energy states are domain walls (kinks),
which, for small values of $g$, are simply characterized by the
presence of nearest-neighbor pairs of antiparallel spins.  Linear
combinations of kink states can be considered as one-particle states
with $O(L^{-1})$ momenta. Hence, there is an infinite number of
excitations with a gap of order $L^{-2}$ and, in the infinite-volume
limit, the ground-state degeneracy is infinite. For the quantum Ising
chain the gap is exactly known for OFBC at $h=0$.  We
have~\cite{CJ-87,CNPV-14,CPV-15}
\begin{equation}
  \Delta_{\rm OFBC}
  = {3g\over 1-g} \, {\pi^2\over L^2} + O(L^{-3}) \,.
  \label{deltaofbc}
\end{equation}
An interesting limiting case is provided by systems with ABC or in the
presence of oppositely-directed boundary fields $\pm h_b$ for a
particular value $h_b = h_{b,c}$ (the behavior for $h\approx h_{b,c}$
shows interesting scaling properties that are discussed in
Refs.~\refcite{CPV-15,CPV-15-iswb}).  In this case, both kink states
and magnetized states are degenerate at the FOQT in the
infinite-volume limit.  The gap still behaves as $L^{-2}$, and, in
particular, we have~\cite{CJ-87,CNPV-14,CPV-15}
\begin{equation}
  \Delta_{\rm ABC} = {g\over 1-g} \, {\pi^2\over L^2} + O(L^{-4}).
  \label{deltaabc}
\end{equation}
In more than one dimensions, this type of behavior can be obtained by
using different BC in the different directions. For instance, one
could use ABC or OFBC in the $x$-direction and PBC or OBC in the
remaining $(d-1)$ directions. This choice would induce domain walls
perpendicular to the $x$ axis.

\begin{figure}[t]
  \begin{center}
    \includegraphics[width=0.5\columnwidth]{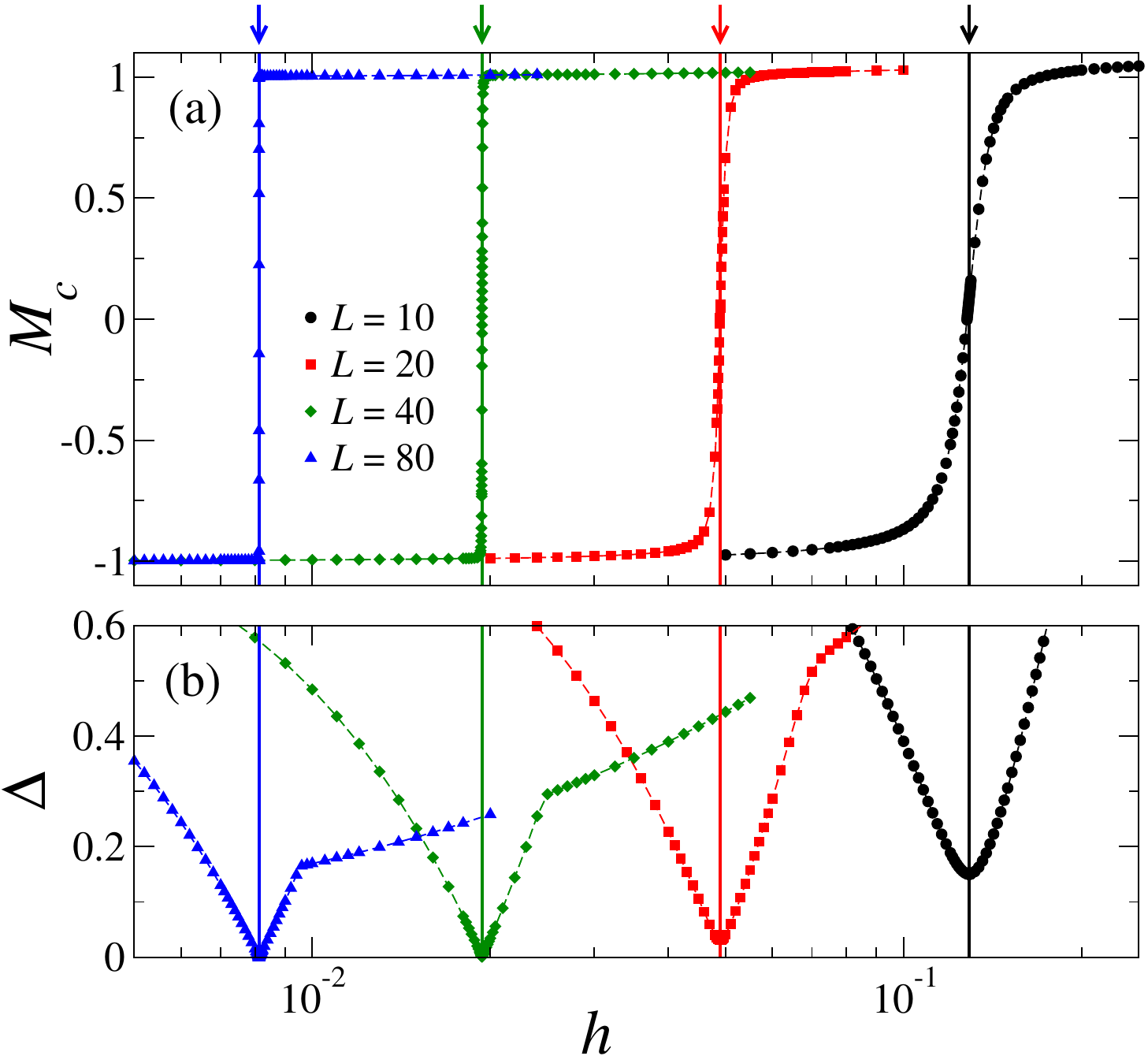}
    \caption{The rescaled local magnetization in the middle of the
      system, $M_c \equiv m_{L/2}/m_0$ [top, panel (a)] and the energy
      gap $\Delta$ [bottom, panel (b)] in the quantum Ising chain with
      EFBC, as a function of the longitudinal field $h$, for a fixed
      transverse field $g=0.8$, and several system sizes.  The
      continuous vertical lines and arrows denote the magnetic fields
      $h_{\rm tr}(L)$ corresponding to the minimum of the energy gap.
      Results from Ref.~\refcite{PRV-18-fowb}.}
    \label{fig:Mc}
  \end{center}
\end{figure}

Finally, we consider nonneutral BC, for instance, equal and fixed BC
(EFBC) which favor one of the two magnetized phases.  They can be
obtained by restricting the Hilbert space to states $|s\rangle$ such
that such that $\hat \sigma^{(1)}_1 |s\rangle = - |s\rangle$ and $\hat
\sigma^{(1)}_{L} |s\rangle = - |s\rangle$ at the boundaries.  The
interplay between the size $L$ and the bulk longitudinal field $h$
turns out to be more complex than that observed with neutral
BC~\cite{PRV-18-fowb}.  In the case of EFBC, for small values of $h$,
observables depend smoothly on $h$, up to $h_{\rm tr}(L)\approx c/L^d$
($c$ is a $g$-dependent constant), where a sharp transition to the
oppositely magnetized phase occurs, see Fig.~\ref{fig:Mc}. The field
$h_{\rm tr}(L)$ can be identified as the value of $h$ where the gap
$\Delta(L,h)$ between the two lowest states is smallest.  The minimum
$\Delta_{\rm min}(L) = \Delta(L,h_{\rm tr}) $ decreases exponentially
with increasing $L$, i.e., $\Delta_{\rm min}(L)\sim e^{-bL}$. Note
that the infinite-volume limit and the limit $h\to 0$ do not
commute. Indeed, the gap $\Delta(L,0)$ is finite for $L\to \infty$.

\subsection{Equilibrium finite-size scaling for neutral boundary conditions}
\label{fssfoqt}

At the FOQTs, the low-energy properties satisfy general FSS laws as a
function of the temperature $T$, the field $h$, and the system size
$L$. We will begin here by discussing the behavior for neutral BC. In
this case, for $h = 0$, the ${\mathbb Z}_2$ symmetry is not broken by
the BC.  If $h\not=0$, the ${\mathbb Z}_2$ transformation maps $h\to
-h$, and all low-energy properties have well defined properties under
$h\to -h$.  For instance, the gap $\Delta$ is even, while the local
magnetization is odd under this transformation.

\subsubsection{General scaling behavior}

To describe the scaling behavior of the low-energy properties as $h$
is turned on, we should identify the appropriate scaling variables.
In a quantum system the relevant low-energy scale is the gap $\Delta$,
i.e., the energy difference of the lowest states, and thus we expect
the interplay between $h$ and the size $L$ to be controlled by the
ratio $\Phi$ between the energy variation $\delta E_h$ due to addition
of the longitudinal field $h$, and the gap $\Delta(L)$ at
$h=0$~\cite{CNPV-14}. Thus, we define the scaling variable
\begin{equation} 
  \Phi \equiv {\delta E_h(L,h)/\Delta(L)} \,,\qquad
  \Delta(L) \equiv \Delta(L,h=0)\,.
  \label{kappah}
\end{equation}
If the BC are such that the lowest-energy states are the magnetized
states, the energy related to the longitudinal field $h$
can be quantified as~\cite{CNPV-14}
\begin{equation}
  \delta E_h(L,h) = 2\, m_0 \,h \,L^d\,,
  \label{deltah}
\end{equation}
for sufficiently small $h$, where $m_0$ is the spontaneous
magnetization, obtained approaching the transition point $h\to 0$
after the infinite-volume limit, see Eq.~\eqref{disco} (of course, the
normalization of $\delta E_h(L,h)$, and in particular the factor two
in its definition, is conventional).  Analogously, the scaling
variable controlling the interplay between $T$ and $L$ is the ratio
\begin{equation}
  \Xi \equiv {T/\Delta(L)} \,.
  \label{kappat}
\end{equation}
The FSS limit corresponds to $L\to \infty$ and $h,T\to 0$, keeping
$\Phi$ and $\Xi$ fixed.  In this limit, the gap $\Delta$ is expected
to behave as~\cite{CNPV-14}
\begin{equation}
  \Delta(L,h) \approx \Delta(L) \, {\cal E}(\Phi) \,.
  \label{fde}
\end{equation}  
By definition ${\cal E}(0)=1$, and ${\cal E}(\Phi)\sim |\Phi|$ for
$\Phi\to\pm \infty$, in order to reproduce the expected linear
behavior $\Delta(L,h) \sim |h|L^d$ for sufficiently large $|h|$.
Analogously, the local and global magnetization are expected to scale
as
\begin{equation}
  m_{\bm x}(L,T,h) \approx m_0 \, {\cal M}_x({\bm x}/L,\Xi,\Phi)\,,\qquad
  M(L,T,h) \approx m_0 \, {\cal M}(\Xi,\Phi)\,.
  \label{efssm}
\end{equation}
Because of the definition (\ref{disco}) of $m_0$, we have ${\cal
  M}(0,\Phi)\to \pm 1$ for $\Phi \to \pm \infty$. Moreover, ${\cal
  M}(\Xi,0)$ vanishes because of the ${\mathbb Z}_2$ invariance.  For
translation-invariant BC (for instance, for PBC), $m_{\bm x}(L,T,h) =
M(L,T,h)$ and ${\cal M}_x({\bm x}/L,\Xi,\Phi)={\cal M}(\Xi,\Phi)$.
The above FSS ans\"atze represent the simplest scaling behaviors
compatible with the discontinuities arising in the thermodynamic
limit.

\begin{figure}[tbp]
  \begin{center}
    \includegraphics[width=0.5\columnwidth]{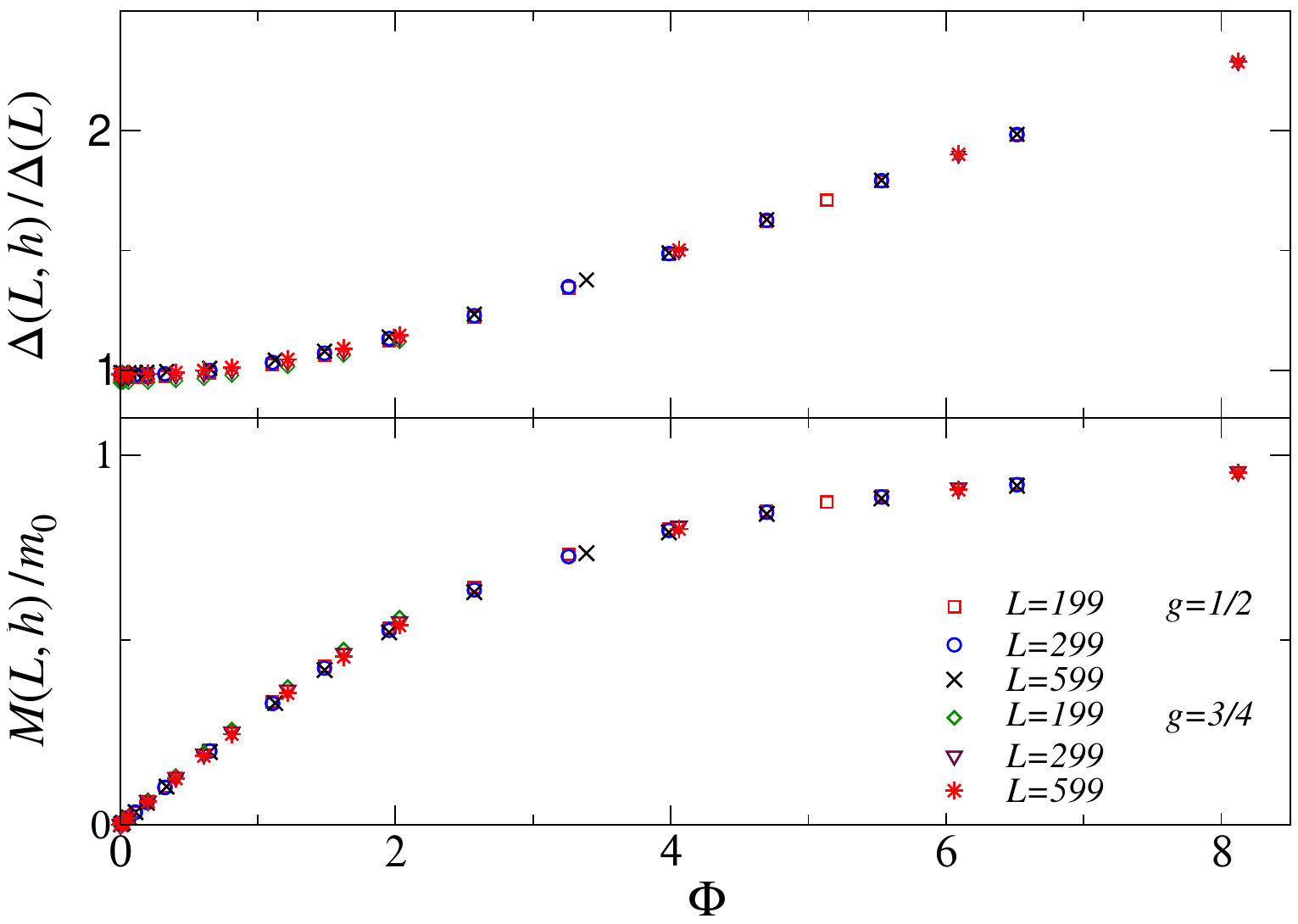}
    \caption{Rescaled energy gap and magnetization for the quantum
      Ising chain with OFBC, at $g=1/2$ and $g=3/4$.  The two panels
      show the ratios $\Delta(L,h)/\Delta(L)$ (top) and $M(L,h)/m_0$
      (bottom) versus $\Phi = 2 m_0 h L/\Delta(L)$ (for OFBC
      $\Delta(L)\sim 1/L^2$). The data approach nontrivial scaling
      curves with increasing $L$, which are independent of $g$,
      supporting Eqs.~\eqref{fde} and~\eqref{efssm}. Results from
      Ref.~\refcite{CNPV-14}. }
    \label{deosx}
  \end{center}
\end{figure}

The previous scaling relations apply generally to any type of neutral
BC, as they simply rely on the fact that the gap is the relevant
energy scale.  Therefore, also in the presence of domain walls, we
expect $\Phi$ and $\Xi$, as defined in Eqs.~(\ref{kappah}) and
(\ref{kappat}), to be the relevant scaling variables. As an example,
in Fig.~\ref{deosx} we report numerical results for the quantum Ising
chain with OFBC for which $\Delta\sim L^{-2}$ at the transition point.
In this case, the lowest-energy states are domain walls, see
Sec.~\ref{gapqutr}. Data satisfy the scaling relations (\ref{efssm})
quite precisely.

We remark that the scaling behavior drastically depends on the nature
of the BC, if one expresses the scaling variables in terms of $L$,
$h$, and $T$, given that the size behavior of $\Delta(L)$ at
criticality depends crucially on the BC.  Indeed, the gap $\Delta$ may
have both an exponential and a power-law dependence on $L$, as
discussed in Sec.~\ref{gapqutr}.  For instance, for the quantum Ising
chain one has: (i)~$\Phi \sim h L e^{cL}$ for OBC; (ii)~$\Phi \sim h
L^{3/2} e^{c L}$ for PBC ($c$ is a nonuniversal positive constant);
(iii)~$\Phi \sim h L^3$ for ABC or OFBC.  This sensitivity to the BC
is one of the main features distinguishing FOQTs and CQTs in
finite-size systems.

It is worth noting that the scaling behaviors (\ref{fde}) and
(\ref{efssm}) in terms of the scaling variables $\Phi$ and $\Xi$
defined in Eqs.~\eqref{kappah} and~\eqref{kappat} also hold at the
CQT, for example at the critical endpoint of the FOQT line.  At the
CQT, the relevant parameters $r=g-g_c$ and $h$ have RG dimensions
$y_r=1/\nu$ and $y_h=(d+z+2-\eta)/2$ (where $\eta$ is the exponent
associated with the power-law decay of the order-parameter two-point
function at the critical point), respectively.  If we fix $r=0$ and
vary the longitudinal field $h$ only, the relevant FSS scaling
variable is $\Phi_c = h\,L^{y_h}$, which, as we show below, can be
identified with the scaling variable $\Phi$ defined in
Eq.~\eqref{kappah}.  Indeed, at the CQT the energy change $\delta E$
due to a perturbation $\hat{H}_h = - h \sum_{\bm x} \hat P_{\bm x}$,
behaves as $\delta E_h(L,h) \sim h \, L^{d - y_p}$ where $y_p$ is the
RG critical dimension of the local operator $\hat P_{\bm x}$ at the RG
fixed point.  If $\hat P_{\bm x} = \hat \sigma^{(1)}_{\bm x}$, $y_p$
satisfies the scaling relation~\cite{CPV-14,RV-21} $y_h + y_p = d +
z$, so that $\delta E_h(L,h) \sim h \, L^{y_h-z}$.  Since, at
criticality, we have $\Delta(L,h=0) \sim L^{-z}$, we obtain $\Phi_c
\sim L^z \delta E_h(L,h) \sim \Phi$.  Analogously, the variable $\Xi$
defined in Eq.~\eqref{kappat} is also the relevant one at the CQT.
Therefore, the FSS relations (\ref{fde}) and (\ref{efssm}) in terms of
the scaling variables~\eqref{kappah} and~\eqref{kappat} hold both to
CQTs and FOQTs.  Differences are encoded in the size dependence of
$\Phi$ and $\Xi$ and in the scaling functions.

The scaling behavior at FOQTs has also been investigated in the
presence of an external local field $h_{\bm x}$~\cite{CNPV-14}. 
The general scaling arguments reported above can also be applied here.  If
the external magnetic field is nonvanishing only at one lattice site
$x$, the energy change is $\delta E\sim 2 m_0 h_x$ and the appropriate
scaling variable is $\Phi_x = 2 m_0 h_x/\Delta(L)$.  Numerical results
for the Ising chain are in full agreement with these scaling
predictions~\cite{CNPV-14}. In the presence of inhomogeneous
conditions, for example when the external fields have a sufficiently
smooth space dependence characterized by a further length scale
  $\ell$, some scaling behaviors emerge as well, involving also the
  spatial position~\cite{CNPV-15-inh}.

\subsubsection{Scaling functions} \label{fsspbcobc}

We now consider a class of neutral BC (for instance, PBC or OBC), such
that the magnetized states represent the lowest-energy excitations.
As discussed in Sec.~\ref{gapqutr}, at the transition 
point $\Delta(L) \sim \exp(-c L^d)$, while
the energy differences $\Delta_n\equiv E_n-E_0$ associated with the
higher excited states ($n>1$) are finite (more generally,
$\Delta_n/\Delta$ decreases exponentially, with possible power
corrections) in the infinite-volume limit.

For sufficiently large $L$, the low-energy properties in the crossover
region $|h|\ll 1$ can be obtained by restricting the theory to the two
lowest-energy states $|0\rangle$ and $|1\rangle$, or equivalently to
the magnetized states $| + \rangle$ and $| -
\rangle$~\cite{CNPV-14,PRV-18}. In this restricted Hilbert space we
can consider an effective Hamiltonian given by
\begin{equation}
  \hat{H}_{2}(h) = \varepsilon \, \hat \sigma^{(3)} + \zeta
  \, \hat \sigma^{(1)}\,.
  \label{hrtds}
\end{equation}
Here $\varepsilon$ corresponds to the energy change due to the
magnetic field $h$, while $\zeta$ is related to the small gap
$\Delta(L)$ at $h=0$~\cite{CNPV-14}, i.e.
\begin{equation}
  \varepsilon =  m_0 h L^d\,,\qquad 
  \zeta = \langle - | \hat H | + \rangle = \tfrac12 
  \Delta(L,h=0) \,.
  \label{deltavarfo}
\end{equation}
The scaling variable $\Phi$ defined in Eq.~(\ref{kappah}) can be
identified as
$\Phi=\varepsilon/\zeta$.  By diagonalizing the effective Hamiltonian
$\hat H_2$, we obtain the ground state
\begin{equation}
| 0 \rangle = \sin(\alpha/2) \, |-\rangle +  
\cos(\alpha/2) \, |+\rangle\,,\qquad \tan \alpha = \Phi^{-1}\,,
\label{eigstatela0}
\end{equation}
and the energy gap $\Delta(L,h)$ as a function of $\varepsilon$ and
$\zeta$.  Moreover, by taking the matrix element of the operator $\hat
\sigma^{(3)}$, i.e.  $\langle 0 | \hat \sigma^{(3)} | 0\rangle$, we
obtain the magnetization of the quantum Ising model.

By matching these results with the scaling equations~\eqref{fde}
and~\eqref{efssm}, one easily obtains the zero-temperature scaling
behaviors~\cite{CNPV-14}  
\begin{equation}
  {\cal E}(\Phi) = \sqrt{1 + \Phi^2}\,, \qquad
  {\cal M}(0,\Phi) =
  {\Phi/\sqrt{1+ \Phi^2}}\,. \label{fdemm}
\end{equation}
Note that, for  OBC the above behavior for the local
magnetization should hold for lattice sites ${\bm x}$ sufficiently far
from the boundaries. For sufficiently small temperatures, i.e., for $T
\sim \Delta$, the scaling function of the magnetization can be
obtained by averaging the magnetization with the Gibbs weight, i.e.,
by considering
\begin{equation}
  M = Z_2^{-1}\, 
  {\rm Tr} \big[ \hat \sigma^{(3)} e^{- \hat H_2/T} \big]\,,\qquad
  Z_2 = {\rm Tr} \big[ e^{-\hat H_2/T}\big] \,.
  \label{twolepa}
\end{equation}

The results reported above have been numerically confirmed for quantum
Ising chains with OBC~\cite{CNPV-14}.  The convergence to the
asymptotic two-level FSS behavior is generally fast, being roughly
controlled by ratio $\Delta/\Delta_n$ between the gap $\Delta \sim
e^{-c L^d}$ and the energy-level differences with the higher states,
i.e. $\Delta_n\equiv E_n-E_0$ for $n>1$, which are finite for
$L\to\infty$. These results supposedly apply to any quantum system, in
which the FOQT is driven by the competition of two different quantum
states that are related by symmetry (the previous expression has been
derived under the assumption that $\langle + | \hat H | + \rangle =
\langle - | \hat H | - \rangle$, but this condition can be easily
relaxed, see Ref.~\refcite{PRV-18-fowb}).  Moreover, the above approach
can be easily extended to systems in which a finite number of states
are quasi degenerate at the FOQT.

When systems at FOQTs develop domain walls, scaling functions cannot
be obtained by performing a two-level truncation, because the
low-energy spectrum at the transition point presents a tower of
excited states (the number of these states becomes infinite for $L\to
\infty$) with $\Delta_{n}=E_n-E_0=O(L^{-2})$. For one-dimensional
chains, it was, however, possible to consider an effective model in
terms of the relevant excitations (of order $L$) and determine the
exact scaling functions, see Refs.~\refcite{CPV-15-iswb,CPV-15} for
details.

\subsection{Equilibrium finite-size scaling for nonneutral boundary
conditions}
\label{fssfebc}

In the case of nonneutral BC (in Ising chains one can consider EFBC),
one should change the definition of the relevant scaling
variables. Indeed, as discussed in Ref.~\refcite{CPV-15-iswb}, no
transition is observed by taking the limit $L\to \infty$ at fixed $h =
0$. For instance, for EFBC the gap $\Delta(L,h=0)$ is finite for $L\to
\infty$. To observe the competition of the two different states, one
should consider the limit $L\to \infty$ along the line $h = h_{\rm
  tr}(L)$ defined in Sec.~\ref{gapqutr}.  The corresponding gap
$\Delta_{\rm min}(L) = \Delta(L,h_{\rm tr})$ vanishes
exponentially. For $h \approx h_{\rm tr}(L)$, a universal FSS behavior
emerges. The relevant scaling variable is
\begin{equation}
  \Phi_{\rm tr} \equiv {2 m_0 L \, [h - h_{\rm tr}(L)]\over
    \Delta_{\rm min}(L)} \, .
  \label{wdef}
\end{equation}
The scaling variable $\Phi_{\rm tr}$ is the analogue of $\Phi$ defined
in Eq.~\eqref{kappah}, with the essential difference that the
finite-size pseudotransition now occurs at $h = h_{\rm tr}(L)$, and
not at $h = 0$.  Therefore, the relevant magnetic energy scale is the
difference between the magnetic energy at $h$ and that at $h_{\rm
  tr}(L)$, while the relevant gap is the one at $h_{\rm tr}(L)$.  For
$h\approx h_{\rm tr}(L)$, observables are expected to satisfy FSS
relations analogous to those reported before
\begin{equation}
  \Delta(L,h) \approx \Delta_{\rm min}(L)\, {\cal E}_f(\Phi_{\rm tr}) \,,
  \qquad m_c(L,h) \approx m_0 \,{\cal M}_{c}(\Phi_{\rm tr}) \,,
  \label{febcsca}
\end{equation}
where $m_c(L,h)$ is the magnetization at the center of the chain. An
analogous relation holds for the average magnetization. These scaling
behaviors have been confirmed by the numerical computations reported
in Ref.~\refcite{PRV-18-fowb}. Since, also in this case, only two states
are relevant for $L\to\infty$, scaling functions can be computed using
a two-level truncation, as discussed in Sec.~\ref{fsspbcobc}. Note
that, at variance with what occurs in the PBC or OBC cases, the energy
differences $\Delta_n$ with the higher states ($n>1$), vanish as a
power of $L$ for $L\to \infty$. Nonetheless, since $\Delta_n/\Delta
\to \infty$ for $L\to \infty$, asymptotically we can still exploit the
two-level truncation. The scaling functions for $\Delta(L,h)$ and
$m_c(L,h)$ are the same as before, apart from a trivial normalization
of the arguments.  For the average magnetization $m(L,h)$ the
expression differs because of the absence of $\mathbb{Z}_2$ symmetry,
see Ref.~\refcite{PRV-18-fowb} for details.

Note that the infinite-volume critical point $h=0$ lies outside the
region in which FSS holds. This implies that, in the definition of the
scaling variable $\Phi_{\rm tr}$, the values of $h_{\rm tr}(L)$ and
$\Delta_{\rm min}(L)$ cannot be replaced with their asymptotic
behaviors.  Analogous behaviors are expected in other FOQTs, for
example in higher dimensions, when BC favor one of the two phases.

\subsection{The quantum Potts model}
\label{potts}

\subsubsection{Definition of the model}

As a second paradigmatic example we consider the one-dimensional
quantum Potts chain~\cite{SP-81,IS-83}, which is the quantum analogue
of the classical two-dimensional Potts
model~\cite{Potts-52,Baxter-73,BN-82,Wu-82}, defined by the classical
Hamiltonian
\begin{equation}
H = - J \sum_{\langle ij\rangle } \delta(s_i,s_j),
\label{cPotts}
\end{equation}
where the sum is over the nearest-neighbor sites of a square lattice,
$s_i$ are spin variables taking $q$ integer values, i.e.,
$s_i=1,...,q$, and $\delta(m,n)=1$ if $m=n$ and zero otherwise.  The
corresponding quantum Hamiltonian is derived from the continuous {\em
  time} limit of the transfer matrix~\cite{SP-81}.  In the quantum
$q$-state Potts chain one considers $q$ states per site and the
Hamiltonian ~\cite{SP-81,IS-83}
\begin{equation}
\hat H = - J_q  \sum_{j=1}^{L-1} \sum_{k=1}^{q-1} 
\Omega_j^k \Omega_{j+1}^{q-k} -
g \sum_{j=1}^L \sum_{k=1}^{q-1} M_j^k 
- h \sum_j \sum_{k=1}^{q} \Omega_j^k\,,
\label{qPotts}
\end{equation}
where $J_q$, $g$, $h$ are the model parameters, and $\Omega_j$ and
$M_j$ are $q\times q$ matrices: $\Omega_{mn} = \delta_{mn} \, e^{i 2
  \pi (n-1)/q}$ and $M_{mn} = \delta_{{\rm mod}(m,q)+1,n}$.  These
matrices commute on different sites and satisfy the
algebra~\cite{SP-81}: $\Omega_j^k \Omega_j^l = \Omega_j^{k+l}$, $M_j^k
M_j^l = M_j^{k+l}$, $ \Omega_j^q = M_j^q = \mathbb{I}$, $M_j^k
\Omega_j^l = \omega^{kl} \Omega_j^l M_i^k$.  For $q=2$ the model is
equivalent to the quantum Ising chain.

For $h=0$ and fixed $J_q$, model (\ref{qPotts}) has a transition at
$g=g_c$ that separates two phases: a disordered phase for $g > g_c$
and an ordered one for $g<g_c$. In the latter phase the system
magnetizes along one of the $q$ {\em directions}.  In the
infinite-volume limit and for $h=0$, the quantum Potts chain satisfies
a self-duality relation~\cite{SP-81} $H(g/J_q) = (g/J_q) H(J_q/g)$.
This implies that the transition point $g_c$ is located at $g_c=J_q$.
In one dimension the transition at $g=g_c$ is continuous for $q \le
4$, and discontinuous for larger values of $q$.  The FOQT at $g_c=J_q$
is somehow qualitatively different from those occurring in quantum
Ising models, in which the energy density is continuous at the
transition, and only the magnetization shows a discontinuity. Indeed,
in the Potts chain both quantities are discontinuous.

The ground-state magnetization 
\begin{equation}
M = \langle \hat\mu_x \rangle\,, \qquad 
\hat\mu_x= {1\over q-1} \left(\sum_{k=1}^{q} \Omega_x^k - 1\right)\,,
\end{equation}
provides an order parameter. For $q>4$, it vanishes in the disordered
phase ($g>g_c$) and, as $g$ is decreased, it jumps discontinuously to
a nonzero value at the FOQT.  In particular, $m_0 = {\rm lim}_{g\to
  g_c^-} \, {\rm lim}_{h\to 0} \, {\rm lim}_{L\to\infty} \, M$ is non
zero for $q>4$~\cite{Kim-81,Baxter-82}. The infinite-volume
ground-state energy density $E$ is also discontinuous across the FOQT.
Indeed, the two limits $E_\pm = {\rm lim}_{g\to g_c^\pm} \, {\rm
  lim}_{L\to\infty} \, E$ differ for $q>4$.  Their difference $\Delta
E \equiv E_+-E_-$ is the analogue of the latent heat in FOCTs.

Beside the transition at $g_c=J_q$, there is also a line of FOQTs
starting at $g_c$ and ending at $g=0$, driven by the magnetic field
$h$ of the Potts Hamiltonian (\ref{qPotts}). These magnetic
transitions are analogous to those discussed in Ising systems.

\subsubsection{Finite-size scaling behavior}
\label{FSSPotts}

As in quantum Ising models, the finite-size behavior of the gap at the
FOQT at $g=g_c$ drastically depends on whether BC are neutral or not.
To specify boundary conditions we define a boundary
Hamiltonian~\cite{IC-99,CNPV-15}
\begin{equation}
\hat{H}_b = - J_q\left[ h_1 \sum_{k=1}^{q-1}
  \Omega_1^k + h_L \sum_{k=1}^{q-1} \Omega_L^k\right],
\end{equation}
which is added to the bulk Hamiltonian (\ref{qPotts}).  OBC and EFBC
correspond to considering $h_1=h_L=0$ and $h_1=h_L=1$.  These two
types of BC are not neutral, as they favor the disordered and ordered
phase, respectively. In both cases, the gap at $g=g_c$ approaches a
constant as $L\to\infty$, so that it is necessary to consider a
pseudotransition coupling $g_{\rm tr}(L)$ to observe a vanishing gap
in the infinite-volume limit (see the discussion in
Sec.~\ref{gapqutr}). Neutral BC are obtained by choosing $h_1=1$ and
$h_L=0$.~\cite{CNPV-15,IC-99} In this case self-duality is preserved
in a finite volume and thus the gap vanishes at $g_c$. The available
numerical results of the gap $\Delta(L)$ for $q=10$ turn out to be
compatible with a power-law behavior~\cite{CNPV-15,CNPV-15-inh,IC-99},
such as $\Delta(L) \sim L^{-\kappa}$ with $\kappa>0$ of order one.

For neutral BC, equilibrium FSS is again controlled by the scaling
variables $\Phi$ and $\Xi$ defined in Eq.~(\ref{kappah}) and
(\ref{kappat}).  Thus, in the case at hand $\Phi = (g-g_c)
L/\Delta(L)$ where $\Delta(L)$ is the gap at the transition. In 
the zero-temperature limit we expect
\begin{eqnarray}
\Delta(L,g) \approx 
\Delta(L)\, {\cal D}(\Phi)\,,\qquad M(L,g) \approx m_0 \,{\cal M}(\Phi)\,.
\label{Pottssca}
\end{eqnarray}
Numerical evidence of the above behaviors was reported in
Ref.~\refcite{CNPV-15}. Similar scalings are expected in the
non-neutral case, simply replacing $\Delta(L)$ with the gap at the
pseudotransition point $g_{\rm tr}(L)$.

The behavior of the model along the line $g<g_c$, where the FOQTs are
driven by the external field $h$, is analogous to that observed for
the quantum Ising model.  For $g < g_c$ and $h=0$, in the
infinite-size limit the ground state is characterized by the
degeneracy of $q$ states.  The degeneracy is lifted in a finite
volume. The behavior of the gap for $h=0$ depends on the nature of the
BC---whether they are neutral or not---and on the nature of the ground
states---whether domain walls are present or not.  In the OBC case,
the energy differences $E_n(h=0,L) - E_0(h=0,L)$ vanish exponentially
as $L\to \infty$ for all $n < q$.  Thus, not only can we predict the
correct scaling variables, but we can also compute the scaling
functions, by performing a truncation of the spectrum: only the lowest
$q$ levels need to be considered.~\cite{CNPV-15} We finally mention
that scaling phenomena at the FOQTs of the Potts model have also been
investigated under inhomogeneous conditions~\cite{CNPV-15-inh}.

\subsection{The ground-state fidelity and the quantum Fisher information}
\label{fidelity}

Quantum transitions in many-body systems are related to significant
qualitative changes of the ground-state and low-excitation properties,
induced by small variations of a driving parameter.  A proper
characterization of their main features may be obtained by using
quantum-information concepts, such as the ground-state fidelity and
the related fidelity susceptibility~\cite{Gu-10,Braun-etal-18,RV-21}.
This quantity plays a central role in quantum estimation theory, as it
is proportional to the quantum Fisher information, which, in turn, is
related to the inverse of the smallest variance achievable in the
estimation of the varying parameter.  The fidelity quantifies the
overlap between the ground states of quantum systems that share the
same Hamiltonian, but with different values of the Hamiltonian
parameters~\cite{Gu-10, Braun-etal-18}.  Its relevance can be traced
back to the Anderson's orthogonality catastrophe~\cite{Anderson-67}:
the overlap of two many-body ground states corresponding to
Hamiltonians differing by a small perturbation vanishes in the
thermodynamic limit.  At quantum transitions the size behavior of the
fidelity significantly differs from that observed in normal
conditions.  Indeed, the fidelity susceptibility is typically much
larger in proximity of quantum transitions than under normal
conditions, so that metrological performances are believed to
drastically improve as a quantum transition is
approached~\cite{ZPC-08, IKCP-08}.

In finite-size quantum Ising models at fixed $g$ and varying $h$, the
ground-state fidelity monitors the changes of the ground-state
normalized wave function $|\Psi_0(L,h)\rangle$ when varying the
control parameter $h$ by $\delta h$. We define the fidelity as
\begin{equation}
  A(L,h,\delta h) \equiv \big| \langle \Psi_0(L,h) | \Psi_0(L,h+\delta h)
  \rangle \big| \,.
\label{fiddef}
\end{equation}
Assuming $\delta h$ small, one can expand Eq.~\eqref{fiddef} in powers
of $\delta h$, obtaining~\cite{Gu-10}
\begin{equation}
  A(L,g,h,\delta h) = 1 - \tfrac12 \delta h^2 \, \chi_A(L,g,h) + 
     O(\delta h^3)\,,
  \label{expfide}
\end{equation}
where $\chi_A$ is the fidelity susceptibility.  The absence of the
linear term in the expansion~\eqref{expfide} is related to the
normalization of the states and is obviously necessary to guarantee
that $A\le 1$.

In normal conditions, for instance, in the disordered phase $g>g_c$,
the fidelity susceptibility is expected to be proportional to the
volume, $\chi_A\sim L^d$.  However, at quantum transitions the
behavior may change.  Indeed, at CQTs there is a nonanalytic
contribution (it is obviously independent of the BC) that behaves as
$\chi_A \approx L^{2y_h} {\cal A}_2(h L^{y_h})$, where $y_h$ is the RG
dimension of the parameter $h$, see, e.g.,
Refs.~\refcite{RV-21,Gu-10,CZ-07} and references therein.  If
$d<2y_h$, as it occurs at the CQT in Ising systems in any dimension,
the singular contribution is the dominant one and the fidelity
susceptibility diverges faster than under normal conditions.  These
results can be trivially extended to any CQT, by replacing $y_h$ with
the RG dimension of the parameter that drives the transition.

At FOQTs, the fidelity susceptibility shows an even stronger
divergence.~\cite{RV-18,RV-21} In terms of the scaling variables
introduced in Sec.~\ref{fssfoqt}, the ground-state fidelity is
conjectured to show the asymptotic FSS behavior
\begin{equation}
  A(L,g,h,\delta h) \approx {\cal A}(\Phi,\delta\Phi) \,,
  \qquad
  \delta\Phi\equiv {\delta E_h(L,\delta_h)/\Delta(L)}\,,
  \label{fiscafo}
\end{equation}
close to the FOQT at $h=0$. This relation
implies
\begin{equation}
 \chi_A(L,h) \approx \left( {\delta\Phi/\delta h}\right)^2 \,
     {\cal A}_2(\Phi) \sim L^{2d} \Delta(L)^{-2} \, {\cal
       A}_2(\Phi)\,.
\label{chifscal}
\end{equation}
Therefore, at FOQTs the fidelity susceptibility shows a divergent
large-$L$ behavior that depends on the BC, as a consequence of the
different possible size behaviors of the gap $\Delta(L)$ discussed in
Sec.~\ref{gapqutr}. In particular, $\chi_A(L,h)$ may scale
exponentially in the size (this is the case of, e.g., PBC or EFBC), or
with a power of $L$. In the ABC case, for instance, since
$\Delta(L)\sim L^{-2}$, the fidelity susceptibility diverges as $L^6$.
As anticipated, these diverging behaviors are much stronger than at
CQTs and under  normal conditions.

\begin{figure}[t]
  \begin{center}
    \includegraphics[width=0.47\columnwidth]{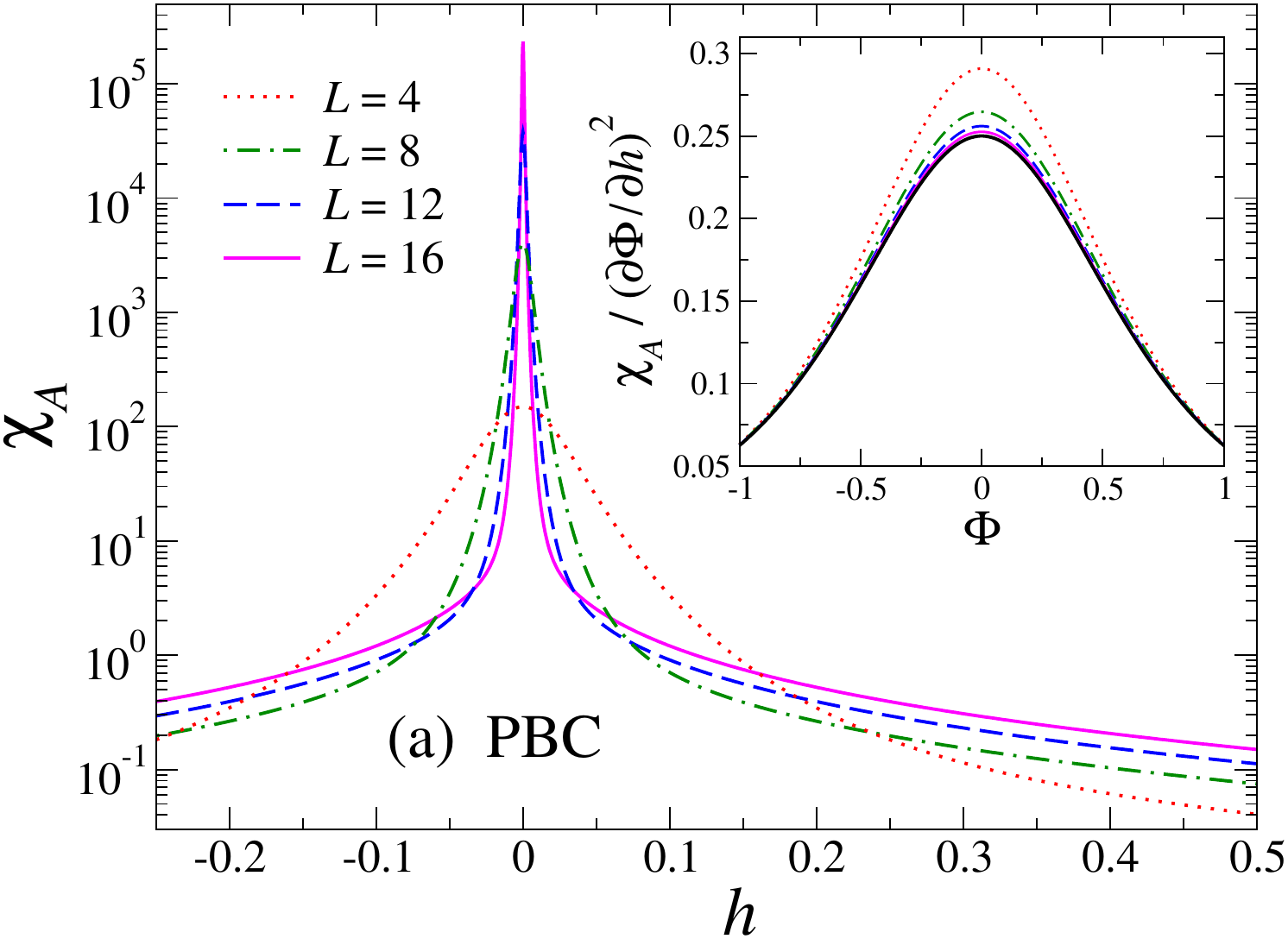}
    \caption{Fidelity susceptibility $\chi_A(L,h)$ for the quantum
      Ising chain with PBC, at fixed $g=0.9$, associated with changes
      of the longitudinal parameter $h$, for some values of $L$.  The
      inset displays $\chi_A/( \partial \Phi / \partial h)^2$, which
      clearly approaches the scaling function ${\cal A}_2(\Phi)$
      (thick black line).  Results from Ref.~\refcite{RV-18}.}
    \label{FidSusc_PBC_g09}
  \end{center}
\end{figure}

If one considers, systems in which the magnetized states are the
relevant low-energy excitations, the scaling function for the fidelity
can be computed by performing a two-level truncation of the states.
For neutral BC such as OBC or PBC, one can use the effective
Hamiltonian ~\eqref{hrtds}, obtaining~\cite{RV-18}
\begin{equation}
  {\cal A}_{2}(\Phi) = {1\over 4 (1 + \Phi^2)^2} \, .
  \label{f2l}
\end{equation}
This expression is expected to hold at any FOQT with only two
symmetric relevant states in the large-$L$ limit. The symmetry
condition is not crucial, and ${\cal A}_{2}(\Phi)$ can also be
computed for nonsymmetric states and therefore for nonneutral BC.

These predictions have been confirmed by numerical computations for
the quantum Ising chain~\cite{RV-18}. Fig.~\ref{FidSusc_PBC_g09} shows
the results for PBC, in which $\Delta(L)$ decreases exponentially.
Eq.~(\ref{fiscafo}) with the scaling function (\ref{f2l}) is clearly
satisfied. The scaling predictions have also been verified for
ABC~\cite{RV-18}, for which $\Phi \approx h L^3$ and therefore
$\chi_A\sim L^{6}$.

\section{Dynamic finite-size scaling at first-order quantum transitions}
\label{foqtdynamics}

In this section we focus on the dynamical behavior of systems close to
FOQTs. We will show that also the dynamic properties obey general FSS
laws that generalize and extend the static FSS relations outlined in
Sec.~\ref{foqt}. As it occurs for equilibrium properties, the
out-of-equilibrium dynamic behavior at FOQTs is very sensitive to the
BC, at variance with what occurs at CQTs, where the power law of the
time scale does not depend on the BC.  We find that the size behavior
of the long-time properties depends on the finite-size gap $\Delta(L)$
at the transition.  If $\Delta(L)$ decreases exponentially---in Ising
systems this occurs when the magnetized states are the lowest-energy
excitations---the relevant time scale of the dynamics increases
exponentially with $L$.  If, instead, $\Delta(L)\sim L^{-\kappa}$, as
it occurs in the presence of domain walls, then the time scale
increases as a power of $L$.

In the following we consider dynamic protocols entailing instantaneous
or slow changes of a Hamiltonian parameter.  Specifically, the
Hamiltonian is supposed to be of the form
\begin{equation}
  \hat{H}[w(t)] = \hat{H}_{0} + w(t) \hat{H}_{p}\,,\qquad
  [\hat{H}_{0},\hat{H}_{p}]\not=0\,.
  \label{timedh}
  \end{equation}
We assume that the Hamiltonian parameters of $\hat{H}_0$ are tuned at
the FOQT, and that $\hat{H}_{p}$ represents a perturbation that takes
the system out of the FOQT.  In the quantum Ising models, $\hat{H}$ is
the Hamiltonian defined in Eq.~\eqref{hisdef}: the Hamiltonian
$\hat{H}_0$ corresponds to $\hat{H}$ with $h=0$ and $g<g_c$, the
parameter $w$ corresponds to the magnetic field $h$ and $\hat{H}_p=-
\sum_{\bm x} \hat \sigma^{(1)}_{\bm x}$.

The equilibrium FSS theory outlined in Sec.~\ref{foqt} can be
straightforwardly extended to the dynamic case by adding the
time-dependent scaling variable
\begin{equation}
  \Theta \equiv \Delta(L)\:t \,.
  \label{thetadeffo}
  \end{equation}
Note that $\Theta$ is also the relevant FSS variable for CQTs.  In the
following we discuss dynamic FSS using the one-dimensional Ising
chain, but these considerations are expected to hold also for more
general systems (without continuous symmetries) and in larger
dimensions.

It is worth mentioning that FOQTs have been studied in the context of
adiabatic quantum computation as well~\cite{AC-09, YKS-10, JLSZ-10,
  LMSS-12}.

\subsection{Quantum quenches at quantum first-order transitions}
\label{neutraldyn}

Instantaneous quench protocols are among the simplest protocols that
can be used to investigate the out-of-equilibrium quantum dynamics of
many-body systems with Hamiltonian (\ref{timedh}). Initially, for $t <
0$, the system is in the ground state of $\hat{H}$ for a specific
value $w_i$ of the parameter $w$, i.e., in the eigenstate
$|\Psi_0(w_i)\rangle$. Then, at $t=0$ one suddenly changes the
Hamiltonian parameter from $w_i$ to $w\neq w_i$, and considers the
quantum evolution defined by the Schr\"odinger equation
\begin{equation}
  i {d |\Psi(t) \rangle \over dt} = \hat H(w) |\Psi(t)\rangle\,,
  \qquad |\Psi(t)\rangle = e^{-i\hat{H}(w)t}|\Psi_0(w_i)\rangle\,.
    \label{schoeq}
  \end{equation}
In the following we consider quantum Ising systems and 
identify the parameters $w_i$ and $w$ with $h_i$ and $h$.

The dynamic FSS relations for the different observables can be
obtained by extending those that hold in the static case.  First, we
introduce the time scaling variable $\Theta$ defined in
Eq.~\eqref{thetadeffo}.  Then, we define the scaling variables
$\Phi_i$ and $\Phi$. They are defined as in Eq.~\eqref{kappah}, using
$h_i$ and $h$, respectively.  The dynamic FSS limit is defined as the
limit $L\to \infty$, keeping the scaling variables $\Phi_i$, $\Phi$
and $\Theta$ fixed.  In this limit, the magnetization behaves as
\begin{equation}
  M(t;L,h_i,h) \approx m_0 \, {\cal M}(\Theta,\Phi_i,\Phi)\,,
  \label{mcheckfoqt}
\end{equation}
where $m_0$ is the spontaneous magnetization defined
in~Eq.~\eqref{sigmasingexp}.  This scaling behavior is expected to
hold for any $g<1$, and the scaling function ${\cal M}$ should be
independent of $g$, apart from trivial normalizations of the
arguments.  As already discussed for the equilibrium FSS in
Sec.~\ref{foqt}, the size dependence of the scaling variables and the
function ${\cal M}$ depend on the BC.

The above dynamic scaling relations can be straightforwardly extended
to any FOQT, by identifying the scaling variable $\Phi$ as the ratio
$\delta E_w(L)/\Delta(L)$, where $\delta E_w(L)$ is the energy
variation associated with the perturbation $w\hat{H}_p$ and
$\Delta(L)$ is the energy difference between the two lowest states at
the transition point. The scaling variable $\Theta$ associated with
time is always defined as in Eq.~\eqref{thetadeffo}.

In the case of neutral BC with magnetized low-energy states, for
instance for OBC or PBC, one may again exploit~\cite{PRV-18} the
two-level truncation of the spectrum discussed in
Sec.~\ref{fsspbcobc}, which holds when $\delta E_w(L,w) = 2 m_0 |w|
L^d = |\Phi| \, \Delta(L) \ll E_2(L)-E_0(L)$ (for systems with PBC or
OBC, $E_2(L)-E_0(L)=O(1)$ in the large-$L$ limit). This allows one to
compute the scaling function ${\cal M}$ associated with the
magnetization, extending the equilibrium computation reported in
Sec.~\ref{fsspbcobc}, see Ref.~\refcite{PRV-18}.  For this purpose, we
consider the unitary evolution of the reduced two-state system with
Hamiltonian~\eqref{hrtds} and define $\varepsilon_i$ and $\varepsilon$
as the values of the parameter corresponding to $h_i$ and $h$.  The
dynamics starts from the two-component ground state corresponding to
the parameter $\varepsilon_i$.  Simple calculations show that, apart
from an irrelevant phase, the time-dependent state $|\Psi_2(t)\rangle$
evolves as~\cite{PRV-18},
\begin{equation}
  |\Psi_2(t)\rangle = \cos\left({\alpha_i-\alpha\over 2}\right) |0\rangle
  + e^{-i \, \Theta \sqrt{1 + \Phi^2}} \sin\left({\alpha_i-\alpha\over 2}\right)
  |1\rangle\,,
  \label{psitfo}
\end{equation}
with $\tan \alpha_i = \Phi_i^{-1}$, $\tan \alpha = \Phi^{-1}$, and
\begin{equation}
  |0\rangle = \sin(\alpha/2) \, |-\rangle +  
  \cos(\alpha/2) \, |+\rangle\,, \qquad 
  |1\rangle =  \cos(\alpha/2) |-\rangle -
  \sin(\alpha/2) \, |+\rangle\,,
\end{equation}
$|\pm \rangle$ being the eigenstates of the operator $\hat \sigma^{(3)}$.
The expectation value $\langle \Psi_2(t) | \hat \sigma^{(3)}
|\Psi_2(t) \rangle$ gives the magnetization, from which
we can infer the dynamic scaling function defined in
Eq.~\eqref{mcheckfoqt}:
\begin{equation}
  {\cal M}(\Theta,\Phi_i,\Phi) = \cos (\alpha-\alpha_i)
  \cos \alpha + \cos \big(\Theta \sqrt{1+\Phi^2}
  \big) \sin(\alpha-\alpha_i) \sin \alpha\,.
  \label{m2lsca}
\end{equation}
Numerical results for the Ising chain with PBC~\cite{PRV-18} confirm
the validity of the scaling Ansatz (\ref{mcheckfoqt}) with the scaling
function (\ref{m2lsca}). The asymptotic FSS behavior is approached
quite rapidly: size corrections turn out to decrease exponentially in
$L$. A similar analisys can be performed for the work fluctuations
associated with the quench protocol~\cite{NRV-19-wo}.  The scaling
predictions obtained by using the two-level truncation are nicely
supported by the numerical results.

The results presented above are expected to be quite general. They
should apply to any FOQT at which there are only two quasi-degenerate
low-energy states which are related by symmetry. For instance, they
should apply to quantum Ising models in any dimension, along their
FOQT line. It is trivial to extend the discussion to FOQT driven by
the competition of two different quantum states that are not related
by symmetry or to systems (Potts models, for instance) in which there
is a finite number of quasi-degenerate states.

We finally mention that, in the case of EFBC which favor a magnetized
phase, see Sec.~\ref{fssfebc}, one should take into account the shift
in the pseudotransition point. As in the static case, the two-level
truncation applies also to dynamic properties, in the appropriate
limit~\cite{PRV-20}.

The dynamic FSS behavior~\eqref{mcheckfoqt} also holds~\cite{PRV-20}
in systems with ABC and OFBC, in which the gap behaves as
$\Delta(L)\sim L^{-2}$. In this case, however, the low-energy behavior
is controlled by a large number (of order $L$) of states, and thus the
dynamics cannot be computed by considering a system with a
finite number of states.  Analogous scaling behaviors are expected in
higher dimensions, for example in the presence of mixed boundaries,
such as ABC or OFBC along one direction, and PBC or OBC along the
others.

We finally mention that out-of-equilibrium scaling behaviors arising
from a quench at FOQTs have also been investigated in more complex
composite models and, in particular, when a central spin is globally
and homogenously coupled to a quantum many-body
system~\cite{RV-19-dec}.

\subsection{Dynamic scaling: Kibble-Zurek dynamics at FOQT}
\label{dynkzfo}

Another interesting out-of-equilibrium dynamics is obtained by slowly
crossing a FOQT. In this case the system parameters, for instance
$w(t)$ in Eq.~(\ref{timedh}), are changed with a very large time
scale $t_s$, starting from an initial value $w_i<0$ on one side of the
FOQT, up to a final $w>0$ on the other side of the transition.  For
instance, in the Kibble-Zurek protocol \cite{Kibble-80,Zurek-96}, at
time $t_i<0$ the system is in the ground state of the Hamiltonian
$H(w_i)$ with $w_i = w(t_i)$, and then it evolves unitarily with
Hamiltonian $H[w(t)]$, where $w(t)= t/t_s$, up to a final time $t_f>0$.
Far from phase transitions, because of the adiabatic theorem, the
system at time $t$ is approximately (exactly in the limit $t_s\to
\infty$) in the ground state of the Hamiltonian $H[w(t)]$.  However,
this adiabatic evolution is not generally realized when the system is
driven across a phase transition, because of the presence of
low-energy modes that have a very long time scale that diverges when
$L\to\infty$.  Thus, in the infinite-volume limit the large-scale
modes do not equilibrate, however large the time scale $t_s$ is. In a
finite volume large-scale modes have a finite time scale $\tau(L)$ and
thus we can observe an out-of-equilibrium scaling behavior due to the
interplay of the two (large) time scales $t_s$ and $\tau(L)$.  Thus, a
slow (quasi-adiabatic) dynamics across critical points allows one to
probe the universal features of the long-range modes emerging at the
transition.

The nature of the out-of-equilibrium FSS behavior depends on the
nature of the transition and on its universality class if it is
continuous, see, e.g., Refs.~\refcite{Kibble-80,Zurek-96,PSSV-11,
  CEGS-12, RV-21, TV-22}.  To be specific, we now consider an Ising
model with $g<g_c$ and a slowly-varying longitudinal field $w(t) =
h(t) = t/t_s$.  At a time $t_i< 0$, the longitudinal field is $h_i =
t_i/t_s$, and the system is in the ground state $|\Psi(t=t_i)\rangle
\equiv |\Psi_0(h_i)\rangle$.  Then, for $t>t_i$, the longitudinal
field varies as $h(t) = t/t_s$ and the system evolves unitarily
according to the Schr\"odinger equation
  \begin{equation}
    {{\rm d} \, |\Psi(t)\rangle \over {\rm d} t} =
    - i \, \hat H[h(t)] \, |\Psi(t)\rangle \,,
    \qquad |\Psi(t=t_i)\rangle = |\Psi_0(h_i)\rangle\,,
    \label{unitdyn}
  \end{equation}
up to a time $t_f > 0$.

In the simultaneous limits $t_s\to\infty$ and $L\to\infty$, the system
shows a universal FSS behavior. To simplify the discussion, we keep
$h_i<0$ fixed in the large-$L$ and large-$t_s$ limit.  Since the
equilibrium FSS behavior should be recovered when $t_s$ is much larger
than the time scale of the low-energy modes, one of the scaling
variables should be identified with the corresponding equilibrium
scaling variable. For neutral BC, the relevant variable is given in
Eq.~\eqref{kappah} and therefore we define
\begin{equation}
  \Phi_{\rm KZ} \equiv {2 m_0 \, h(t) \, L^d \over \Delta(L)} = {2 m_0 t L^d
    \over \Delta(L) \: t_s}\,.
  \label{katdef}
\end{equation}
The second scaling variable is $\Theta=\Delta(L) \,t$ 
defined in Eq.~(\ref{thetadeffo}). Equivalently, we can define
\begin{equation}
  \Upsilon \equiv {\Theta\over \Phi_{\rm KZ}} =
           {\Delta^2(L) \: t_s\over 2 m_0 L^d}\,, 
  \label{taudef}
\end{equation}
which is independent of the time $t$.  The dynamic FSS limit
corresponds to $t, t_s, L\,\to\infty$, keeping the scaling variables
$\Phi_{\rm KZ}$ and $\Upsilon$ fixed. In this limit the magnetization
is expected to scale as
\begin{equation}
  M(t,t_s,h_i,L) \approx m_0\,{\cal M}(\Upsilon, \Phi_{\rm KZ}) \,,
  \label{mtsl}
\end{equation}
independently of $h_i<0$.  In the adiabatic limit ($t,t_s\to \infty$
at fixed $L$ and $t/t_s$), ${\cal M}(\Upsilon\to\infty, \Phi_{\rm
  KZ})$ must reproduce the equilibrium FSS function~\eqref{efssm}. The
scaling functions are expected to be universal (i.e., independent of
$g$ along the FOQT line, for a given class of BC).  Note that the
dynamic FSS behavior develops in a narrow range of magnetic fields.
Indeed, since $\Phi_{\rm KZ}$ is kept fixed in the dynamic FSS limit,
the relevant scaling behavior develops in the interval $|h| \lesssim
\Delta(L)/L^d$, which rapidly shrinks when increasing $L$.  This implies
that the dynamic FSS behavior is independent of the initial value
$h_i<0$ if it is kept fixed in the large-$L$ limit.

The above scaling behaviors are expected to hold for generic neutral
BC, both in the presence of magnetized ground states and in the
presence of domain walls.  They can also be extended to nonneutral BC,
by replacing \cite{PRV-20} $h(t)$ with $h(t) - h_{\rm tr}(L)$, where
$h_{\rm tr}(L)$ is defined in Sec.~\ref{gapqutr}.  The previous
scaling expressions are quite general and can be straightforwardly
extended to any FOQT.  The previous dynamic FSS relations have been
numerically verified in Refs.~\refcite{PRV-18-def, PRV-20} for Ising
chains with global and local time-dependent perturbations and for
several types of BC.

If we consider neutral BC with magnetized ground states, we can also
compute the dynamic scaling functions associated with the Kibble-Zurek
protocol. Indeed, since the low-energy behavior is controlled by the
two lowest-energy states, we can consider the effective
Hamiltonian~(\ref{hrtds}), which now becomes time-dependent:
\begin{equation}
  \hat{H}_{r} = \varepsilon(t) \, \hat \sigma^{(3)} + \zeta \,
  \hat \sigma^{(1)}\,,\qquad \varepsilon(t) = {m_0 \, h(t) \, L^d}
  = m_0 t L^d / t_s\,,\quad \zeta = \Delta(L) / 2\,.
  \label{hrdef2t}
\end{equation}
The system is thus equivalent to a two-level quantum mechanical system
in which the energy separation of the two levels is a linear function
of time. Its dynamics was first investigated by Landau and
Zener~\cite{Landau-32, Zener-32} and solved exactly in
Ref.~\refcite{VG-96}.  If $|\Psi_2(t)\rangle$ is the solution of the
Schr\"odinger equation with the initial condition $|\Psi_2(t_i)\rangle
= |+\rangle$ (where $|+\rangle$ is the eigenstate of $\hat
\sigma^{(3)}$ with positive eigenvalue), the dynamic FSS function of
the magnetization can be computed, obtaining
~\cite{PRV-18-def, PRV-20}
\begin{equation}
  {\cal M}(\Upsilon, \Phi_{\rm KZ}) = \langle \Psi_2(t)
  | \hat \sigma^{(3)}|\Psi_2(t)\rangle =
  -1 + \tfrac12 \Upsilon e^{-{\pi
      \Upsilon\over 8}} \left| D_{-1+i{\Upsilon\over 4}} (e^{i{3\pi\over
      4}}\sqrt{\Upsilon} \,\Phi_{\rm KZ}) \right|^2 \,, 
  \label{fsigmasol}
\end{equation}
where $D_\nu(x)$ is the parabolic cylinder function~\cite{AS-1964}.
In this case, the deviations from the asymptotic dynamic FSS behavior
are controlled by the ratio between $\Delta(L)\sim e^{-cL^d}$ and
$\Delta_2=O(1)$.  Therefore, the approach to the asymptotic behavior
is expected to be exponentially fast~\cite{PRV-18-def}.

The above results are expected to extend to generic FOQTs in which the
universal features of the dynamics of the low-energy modes are
effectively encoded in the dynamics of a two-level system.  It is
worth mentioning that an analogous dynamic scaling behavior is
observed by considering time-dependent localized defects, such as a
time-dependent magnetic field localized in one site of an Ising chain,
essentially because the chain behaves rigidly along the FOQT line.

We point out that the dynamic scaling behaviors that have been
outlined above have been numerically observed in relatively small
systems.  Therefore, given the need for high accuracy without
necessarily reaching scalability to large sizes, the available
technology for probing the coherent quantum dynamics of interacting
systems, using, for instance, ultracold atoms in optical
lattices~\cite{Bloch-08, Simon-etal-11}, trapped
ions~\cite{Edwards-etal-10, Islam-etal-11, LMD-11, Kim-etal-11,
  Richerme-etal-14, Jurcevic-etal-14, Debnath-etal-16}, as well as
Rydberg atoms in arrays of optical microtraps~\cite{Labuhn-etal-16,
  Keesling-etal-19}, could offer possible playgrounds where the
envisioned behaviors at FOQTs can be observed.

Metastability phenomena in Kibble-Zurek protocols in quantum Ising
chains have also been investigated in Ref.~\refcite{SCD-21}, evidentiating
quantized nucleation mechanisms.  The out-of-equilibrium dynamics at
FOQTs has been also investigated in the presence of
dissipation~\cite{DRV-20,NF-20} in the Lindblad
framework~\cite{Lindblad-76,GKS-76,BP-book,RH-book}.

\section{Quantum-to-classical mapping}
\label{qtoc}

In the previous sections we have discussed FSS in quantum
systems. However, historically, the basic concepts underlying our
present understanding of critical and first-order transitions, were
developed in a classical setting, in several seminal works by
Kadanoff, Fisher, Wilson, among others (see, e.g.,
Refs.~\refcite{Fisher-71, WK-74, Fisher-74, Ma-book, BGZ-76,
  Wegner-76, Aharony-76, Wilson-83}).  Their extension to quantum
systems is based on the quantum-to-classical mapping, which allows one
to map a quantum system defined in a spatial volume $V_s$ onto a
classical one defined in a box of volume $V_c = V_s \times L_T$, with
$L_T=1/T$ (using the appropriate units)~\cite{SGCS-97, Sachdev-book,
  CPV-14}. Under the quantum-to-classical mapping, the inverse
temperature corresponds to the system size in an imaginary time
direction.  The BC along the imaginary time are periodic or
antiperiodic for bosonic and fermionic excitations, respectively.
Thus, the universal FSS behavior observed at a quantum transition in
$d$ dimensions is analogous to the FSS behavior in a corresponding
$D$-dimensional classical system with $D=d+1$.

It is important to remark that the quantum-to-classical mapping does
not generally lead to standard classical isotropic systems in thermal
equilibrium. First, in many interesting instances one obtains
complex-valued Boltzmann weights (a well-known example is the Berry
phase).  Moreover, the corresponding classical systems are generally
anisotropic.  If the dynamic exponent $z$ that controls the size behavior 
of the gap is equal to 1, as for the Ising CQT,
the anisotropy is weak, as in the classical Ising model with
direction-dependent couplings. In these cases, a straightforward
rescaling of the imaginary time allows one to recover space-time
rotationally invariant (relativistic) $\Phi^4$ statistical field
theories.  However, there are also interesting transitions in which
$z\not=1$, such as the superfluid-to-vacuum and Mott transitions of
lattice particle systems described by the Hubbard and Bose-Hubbard
models \cite{FWGF-89,Sachdev-book}. In this case, we have $z=2$ when
the transitions are driven by the chemical potential.  For CQTs with
$z \neq 1$, the anisotropy is strong, i.e., correlations have
different exponents in the spatial and thermal directions.  Indeed, in
the case of quantum systems of size $L$, under a RG rescaling by a
factor $b$ such that $\xi\to\xi/b$ and $L\to L/b$, the additional
dimension related to the temperature rescales differently,
as $L_T \to L_T/b^z$. However, a FSS theory has also been
established in classical strongly anisotropic
systems~\cite{Zia-review}.

\section{Equilibrium finite-size scaling at classical first-order transitions}
\label{sec5}

\subsection{General scaling results in discrete spin systems}
\label{sec5.1}

FSS for first-order classical transitions (FOCTs) was established in a series of
seminal papers~\cite{NN-75,FB-82,PF-83,BN-82}, generalizing the ideas
that had been previously developed in the context of critical
transitions.  Michael Fisher was activily involved in the development
of a general FSS theory for first-order transitions. We do not plan to
review here these very well-known issues, see, e.g.,
Ref.~\refcite{Binder-87}.  We will only present some results, due to
Michael Fisher and collaborators, that have a direct counterpart in
quantum systems and directly relate to what we have been discussing in
the quantum setting.

Ref.~\refcite{PF-83} studied FSS for generic Ising systems with PBC,
considering asymmetric $D$-dimensional classical systems of size
$L^{d}\times L_\parallel$ with $D=d+1$.  For ``block" geometries in
which the FSS limit is taken at fixed finite and nonvanishing ratio
$L/L_\parallel$, the basic scaling variable turns out to be the
dimensionless ratio of the total bulk ordering energy and the thermal
energy. In Ising systems, along the low-temperature first-order
transition line, the natural variable is therefore
\begin{equation}
      \Phi_b = m_0 h V/T\,, \qquad V = L^{d} L_\parallel=L^D\times
      (L_\parallel/L)\,,
\label{Phib-classical}
\end{equation}
where $T$ is the temperature, $h$ the magnetic field, and $m_0$ the
spontaneous magnetization.  The free energy and the magnetization
should therefore scale as
\begin{equation}
  F(h,T) = A(T) \, {\cal F}(\Phi_b)\,,\qquad m(h,T) = m_0(T) \,
  {\cal M}(\Phi_b)\,,
\end{equation}
a result which was then proved rigorously, under very general
conditions, in Ref.~\refcite{BK-90}.  It was also noted that this
behavior does not hold when $L_\parallel\gg L$, the geometry that is
relevant for quantum systems ($L_\parallel$ is the size of the
imaginary-time direction that appears in the quantum-to-classical
mapping). In this case, fluctuations break the system into successive
regions of up and down magnetization with characteristic length
\begin{equation}
\xi_\parallel(T,L) \sim \exp[\sigma(T) L^{d}], 
\end{equation}
where $\sigma(T)$ is the reduced interfacial tension. Correspondingly,
the relevant scaling variables are 
\begin{equation}
    \Phi = m_0 h L^{d} \xi_\parallel(T,L)/T\,, \qquad
    \Xi  = \xi_\parallel(T,L)/L_\parallel\, .
\label{Phi-cilindro-classico}
\end{equation}
Under the quantum-to-classical mapping the $D$-dimensional classical
system with $L_\parallel\gg L$ is mapped onto a $d$-dimensional quantum
sistem, $\xi_\parallel(T,L)/T$ and $1/L_\parallel$ playing the role of the
energy gap and of the temperature $T$ of the quantum system,
respectively.  Thus, for PBC, Fisher and Privman had already
identified the variables~(\ref{kappah}) and (\ref{kappat}) as the
correct scaling variables for FOQTs. Moreover, they postulated what
they called "the two-eigenvalue dominance": Scaling functions can be
computed by considering only the two largest eigenvalues of the
transfer matrix. In the quantum context this corresponds to the
two-level truncation discussed in Sec.~\ref{fsspbcobc}.  Obviously,
the quantum-to-classical mapping can also be used to extend quantum
results to classical systems, i.e. we can use the results of
Sec.~\ref{fssfoqt} to predict the scaling behavior in cylinder-like
systems with PBC in the {\em long} direction and different types of BC
in the other ones. For instance, the results for the quantum Ising
chain with ABC immediately predict the scaling behavior of a classical
Ising model in a two-dimensional strip with ABC along the parallel
boundaries at finite distance $L$. In this case we would predict that
the magnetization scales as $M(L,h) \approx m_0 {\cal M} (hL^3)$,
where $L$ is the size of the strip. The same result would be obtained
by considering sufficiently strong oppositely-oriented boundary
fields.  For nonneutral BC, the quantum analysis predicts a
surface-induced shift of the pseudotransition $h_{\rm tr}(L)$ (note
that, by applying the quantum-to-classical mapping, we expect $h_{\rm
  tr}(L)\sim L^{-d}$), a well-known phenomenon in the classical
case.~\cite{FB-82,PR-90} Moreover, in the definition of the scaling
variables $h$ should be replaced by $h - h_{\rm tr}(L)$, as verified
explicitly in classical systems~\cite{PR-90}.

We also mention that the effects of spatially inhomogeneous
conditions, such as a temperature gradient, have been also analyzed at
FOCTs,~\cite{BDV-14} showing the emergence of peculiar scaling
behaviors, similar to those observed at FOQTs.

\subsection{General scaling results in continuous spin systems} 
\label{sec5.2}

The analysis of the previous section can be extended to
$D$-dimensional magnetic systems with a continuous O$(N)$
symmetry. Results for PBC were obtained Fisher and Privman in
Ref.~\refcite{FP-85}. Again, the behavior depends on the geometry of the
system.  In cubic systems the modulus of the magnetization is
essentially constant, while its direction is randomly distributed in
the sphere, to recover the O($N$) invariance. The relevant scaling
variable is $\Phi_b$ defined in Eq.~(\ref{Phib-classical}).
Corrections, such as those arising from spin-wave modes, are expected
to decay as $1/L$~\cite{FP-85}.

FSS substantially changes in anisotropic geometries. If we consider
systems of size $L^{d} \times L_\parallel$ and $L_\parallel \gg L$, fluctuations
give rise to spin waves in the ``long" direction, with a finite length
scale~\cite{FP-85} $\xi_\parallel$ that scales as
\begin{equation}
\xi_\parallel \approx L_\parallel \, f_p(L_\parallel/L^2),
\label{xiparsca}
\end{equation}
where $f_p(x)$ is a scaling function.  The relevant scaling variables
are again $\Phi$ and $\Xi$ given in
Eq.~(\ref{Phi-cilindro-classico}). These predictions have been
confirmed by analytical and numerical computations for the three-dimensional
O($N$) vector model~\cite{PV-16,SW-18}.

We can use the above-reported results and the quantum-to-classical
mapping to conclude that the variables~(\ref{kappah}) and
(\ref{kappat}) are the relevant FSS variables also for a quantum
system with continuous symmetry (at least when PBC are used).  This
result provides additional evidence that the only relevant system
variable for FSS is the finite-size gap $\Delta(L)$ at the transition
(or at the pseudocritical transition, for nonneutral boundary
conditions).  Moreover, Eq.~(\ref{xiparsca}) implies that the gap
scales as $1/L^2$, so that $\Phi \sim h L^{d+2}$ for a quantum
$d$-dimensional system with continuous symmetry.

\subsection{Finite-size behavior of cumulants at 
first-order transitions}

First-order transitions are characterized by the discontinuity of
thermodynamic quantities. In $D$-dimensional finite systems they are
smeared out, but they still give rise to easily identifiable
properties. For instance, the specific heat and the magnetic
susceptibility show a sharp maximum that diverges as a power of the
volume and the same is true for the cumulants of the order parameter
in magnetic transitions.

A FSS analysis of the crossover behavior in the coexistence region
relies on some phenomenological approximations for the distribution of
the quantity one is considering
\cite{CLB-86,Binder-87,BVDRSL-92,VRSB-93}. Let us consider a generic
intensive scalar quantity $A$ that is discontinuous at a transition
that separates a single stable phase from $q$ equivalent phases.  For
the magnetic low-temperature Ising transitions (or the equivalent
nonsymmetric liquid-vapor transitions), $q$ is 1. Instead, thermal
first-order transitions in Ising-like or Potts models have $q\ge 2$.
The basic assumption is that, close to the transition, $A$ has a
bimodal distribution,
\begin{equation}
P(A) = {w L^{D/2} \over \sqrt{2 \pi B_+} }
    \exp \left(- {(A - A_+)^2 L^D\over 2 B_+} \right) + 
      {(1-w) L^{D/2} \over \sqrt{2 \pi B_-} }
    \exp \left(- {(A - A_-)^2 L^D\over 2 B_-} \right)\,,
\label{distributionP}
\end{equation}
where $A_\pm$ and $B_\pm$ are regular functions of the parameter
driving the transition. This dependence is irrelevant for the
discussion below, and thus $A_\pm$ and $B_\pm$ will represent the
values of the corresponding functions at the transition point. We
assume the "+" phase to be unique, while there are $q$ equivalent
"$-$" phases. The parameter $w$ is related to the weight of the
different phases.  Since $\langle A \rangle = w A_+ + (1-w) A_-$, we
immediately see that
\begin{equation}
     {w\over 1-w} = {1\over q} \exp(-\Delta F/T)\,,
\end{equation}
where $\Delta F$ is the free energy difference between the "+" phase
and one of the "$-$" phases.~\footnote{Note that the expression
(\ref{distributionP}) differs from that reported in
Ref.~\refcite{CLB-86}, where the relative weights of the two Gaussians
were computed using an equal-height prescription, but agrees with what
reported in Ref.~\refcite{Binder-87} (see Sec.~3 of
Ref.~\refcite{BK-90} for a rigorous discussion).}  At the transition
point $\Delta F = 0$ and therefore $w=1/(q+1)$: the average value
$\langle A \rangle$ is obtained as the average of $A$ over the $(q+1)$
coexisting phases. If the transition is driven by varying a parameter
$\lambda$ coupled to $A$ (i.e., the Hamiltonian contains a term
$\lambda \sum_x A_x$), then we have $\Delta F = (\lambda - \lambda_c)
(A_+ - A_-) L^D$, close to the transition point $\lambda = \lambda_c$,
so that
\begin{equation}
w = {e^{-\beta A_+ L^D \Delta \lambda} \over e^{-\beta A_+ L^D \Delta
    \lambda} + q e^{-\beta A_- L^D \Delta \lambda} }\,,\qquad \Delta
\lambda = \lambda - \lambda_c\,.
\end{equation}

The distribution (\ref{distributionP}) is motivated by
phenomenological considerations, although many of the results we will
summarize below have been proved rigorously for systems with
PBC~\cite{BI-89,BK-90}.  It is important to stress that the Ansatz
(\ref{distributionP}) is not expected to be true for any type of
system geometry and boundary conditions.  For instance, it should not
hold for very elongated geometries or for systems with mixed boundary
conditions that favor different phases, since, in these cases, stable
domain walls are present in the system.  In general, we expect the
behavior to hold whenever interfaces have an exponentially small
probability.

Distribution (\ref{distributionP}) allows us to predict the behavior of 
cumulants of $A$ in the coexistence region. We first consider the 
susceptibility (if $A$ is the energy, $\chi_A$ is proportional to the 
specific heat)
\begin{equation}
\chi_A = L^D (\langle A^2 \rangle - \langle A \rangle^2)\,,
\end{equation}
which takes the values $B_+$ and $B_-$ in the two phases. In the coexistence
region, $\chi_A$ has a sharp maximum with 
\begin{equation}
\chi_{A,\rm max} =  {1\over 4} (A_+ - A_-)^2 L^D [1 + O(L^{-D})]\,.
\end{equation}
For nonsymmetric systems, the maximum does not occur at the critical
point, but is shifted by a quantity proportional to $L^{-2D}$ for
$q=1$ and to $L^{-D}$ for $q\ge 2$ (see Ref.~\refcite{BK-90} for a
critical discussion and proofs in the PBC case).

It is also interesting to consider the Binder parameter 
\begin{equation}
   U_A = {\langle A^4 \rangle \over \langle A^2 \rangle^2}\,.
\label{defUA}
\end{equation}
The behavior of $U_A$ depends on the nature of the quantity that is 
considered.  If both $A_+$ and $A_-$ are non vanishing at the
transition, then $U_A$ takes the value 1 at both sides of the
transition and shows a maximum in between (the position of the maximum
converges to the transition point with corrections of order $L^{-D}$)
with
\begin{equation}
   U_{A,\rm max} = {(A_+^2 + A_-^2)^2 \over 4 A_+^2 A_-^2} + O (L^{-D})\,.
\end{equation}
At the transition point 
the parameter is not trivial:
It takes the value
\begin{equation}
   U_{A,c} = {(q+1) (A_+^4 + q A_-^4) \over (A_+^2 + q A_-^2)^2} + O (L^{-D})\,.
   \label{Uac}
\end{equation}
This is the expected generic behavior at thermal transitions where $A$
is the energy density.  In the Ising model along the low-temperature
transition line, if $A$ is the magnetization $M$, then $A_+ = - A_- =
m_0$, where $m_0$ is the spontaneous magnetization. In this case, the
Binder parameter behaves trivially as $U_m$ is 1 in the whole
coexistence region (it is trivial to verify that $U_{m,\rm max} = 1$).

The behavior drastically changes if $A_+ = 0$ or $A_-=0$.  Assuming
for definiteness that $A_+ = 0$, $U_A$ varies from 3 (in the ``+"
phase) to 1 and has a maximum for $w \approx 1 - B_+
L^{-D}/A_-^2$---therefore in the ``+" phase---with
\begin{equation} 
U_{A,\rm max} = {A_-^2 \over 4 B_+} L^D\,.
\label{UAmax-div}
\end{equation}
At the transition point $w=1/(q+1)$, we have instead $U_{A,c} =
(1+q)/q + O(L^{-D})$.  A scalar order parameter at a thermal
FOCT is expected to have this type of behavior,
since the its average vanishes in the disordered "+" phase.  The same
qualitative behavior is expected in the Potts models defined by the
Hamiltonian (\ref{cPotts}), although, in this case, one should take
into account the multicomponent nature of the order parameter
\cite{BVDRSL-92,VRSB-93}.

It should be noted that, in many cases, the definition (\ref{defUA})
is somewhat arbitrary, as $A$ may be defined modulo an additive
constant (this is obviously the case if $A$ is the energy). It is thus
natural to define
\begin{equation}
  U_A(A_0) = {\langle (A - A_0)^4 \rangle \over \langle (A - A_0)^2 \rangle^2}\,,
\label{UAsubtr}
\end{equation}
where $A_0$ is an arbitrary constant. The maximum $U_{A,\rm max}(A_0)$
for a given value of $A_0$ shows a peak that becomes sharper and
sharper as $A_0$ gets closer to $A_+$ or $A_-$. In particular the
maximum diverges as $L^D$, when $A_0$ is equal to $A_+$ or $A_-$.

The previous results can be used to distinguish first-order
transitions from continuous ones.  At FOCTs the probability
distributions of the energy and of the magnetization are expected to
show a double peak for large values of $L$.  Therefore, two peaks in
the distributions are often taken as an indication of a first-order
transition.  However, as discussed, e.g., in
Refs.~\refcite{Ape-90,FMOU-90,Mc-90,Billoire-95} and references therein,
the observation of two maxima in the distribution of the energy is not
sufficient to conclude that the transition is a FOCT.  For
instance, in the two-dimensional Potts model with $q=3$ and
$q=4$~\cite{FMOU-90,Mc-90}, double-peak distributions are observed,
even if the transition is known to be continuous.  Analogously, in the
3D Ising model the distribution of the magnetization at the critical
point has two maxima \cite{TB-00}.  In order to identify definitely a
first-order transition, it is necessary to perform a more careful
analysis of the large-$L$ scaling behavior of the distributions.

A possible approach consists in considering the large-$L$ behavior of
the specific heat or of the susceptibility of the order parameter,
that should have a maximum that diverges as
$L^D$.~\cite{Ape-90,FMOU-90,Mc-90,Billoire-95} In practice, one should
compute the maximum $\chi_{A,\rm max}(L)$ for each value of $L$ and
then fit the results to a power law $a L^p$. If one obtains $p\approx
D$, the transition is of first order; otherwise, the transition is
continuous, $p$ being related to standard critical exponents. If $A$
is the energy density, then $\chi_A$ is the specific heat that diverges as
$L^{\alpha/\nu}$ (if $\alpha$ is positive), while if $A$ is the
magnetization, then $\chi_A\sim L^{2-\eta}$. This approach works
nicely for strongly discontinuous transitions. In the opposite case,
however, the asymptotic behavior $\chi_A\sim L^D$ may set in for
values of $L$ that are much larger than those considered in the
simulations. Thus, data may show an effective scaling $\chi_{A,\rm
  max}(L) \sim L^p$, with $p$ significantly smaller than $D$,
effectively mimicking a continuous transition (the 3-state Potts model
in three dimensions is a good example, see
Refs.~\refcite{Ape-90,FMOU-90,Mc-90,Billoire-95}). At order-disorder
magnetic transitions, the Binder cumulant of the magnetization usually
provides a better indicator~\cite{VRSB-93}.  Indeed, it diverges at
first-order transitions, see Eq.~(\ref{UAmax-div}), while it is smooth
and finite at continuous transitions. Thus, the observation that
$U_{m,\rm max}(L)$ increases with $L$ is an evidence of the
discontinuous nature of the transition, even if the maximum does not
scale as $L^D$, as it should do asymptotically.  This idea have been
exploited to determine the nature of the transitions in several
models, including systems with gauge symmetries, such as the Abelian
Higgs model and scalar chromodynamics, see, e.g.,
Refs.~\refcite{PV-19,BPV-21,BPV-19,BFPV-21}.

Finally, note that we can exploit the behavior of the Binder parameter
to estimate the transition point also at first-order transitions. For
instance, at a thermal transition, the temperature $T_{\rm max}(L)$,
where $U_A(L)$ attains its maximum, converges to $T_c$.  One could
also use the crossing method \cite{Binder-81}.  Since the value
$U_{A,c}$ that the Binder parameter takes at the transition,
Eq.~(\ref{Uac}), differs from the value it takes outside the
coexistence region, the curves $U_A(L)$ for different values of $L$
have an intersection point that converges to $T_c$ as $L$ becomes
large.  The energy cumulant $U_E$ can also be used~\cite{CLB-86}. At
continuous transitions, $U_E$ is asymptotically approaching 1, while
a peak should be observed at first-order transitions. Note that it is
probably useful to consider the subtracted cumulant defined in
Eq.~(\ref{UAsubtr}) with $E_0$ close to the value that the energy
takes in the transition region, as this would provide sharper maxima.

It is possible to generalize the previous discussion to systems with
continuous symmetries. For the energy, nothing changes: all results
hold also in this case, see, e.g., Ref.~\refcite{CPPV-04}.  For the
order parameter, one should take into account its multicomponent
nature, which is only expected to give rise to quantitative changes
with respect to the previous results.

\section{Dynamic scaling at classical first-order transitions}
\label{sec6}

As we have done for the quantum case, we will present here some
general results for the equilibrium and out-of-equilibrium dynamics of
classical systems, focusing again on the behavior close to FOCTs
(see, e.g., Refs.~\refcite{HH-77,Enz-79,CG-05,FM-06} for
reviews on the critical dynamics at continuous transitions).
At FOCTs several interesting dynamic phenomena occur. Here we will
only consider some specific topics that can be somehow related to what
we have discussed in the quantum case.

\subsection{Equilibrium dynamics}

Let us consider the dynamic behavior of a $D$-dimensional system in
equilibrium, close to a FOCT, focusing on the long-time behavior of
global observables.  As it occurs at critical transitions
\cite{HH-77,FM-06}, scaling variables and scaling functions depend on
some general features of the dynamics.  The results presented here
refer to the purely dissipative dynamics~\cite{Ma-book,HH-77}, driven
by one dissipative coupling without conservation laws (model A of the
classification reported in Ref.~\refcite{HH-77}), which can be easily
realized in lattice models by using the Metropolis or the heat-bath
algorithm~\cite{Creutz-book}.  As we shall see, in a finite
volume the FSS relations turn out to crucially depend on the BC and on
the geometry of the system, similarly to what happens at FOQTs.

As in the quantum case, the extension of the FSS theory to the dynamics is
straightforward. Beside the scaling variable $\Phi$ defined in
Sec.~\ref{sec5}, it is enough to consider the additional scaling
variable $\Theta = t/\tau(L)$, where $\tau(L)$ represents the time
scale of the slowest modes of the system.  In the dynamic FSS limit,
i.e., at fixed $\Phi$ and $\Theta$, the long-time behavior of the
system shows a universal scaling behavior.  For instance, in the
dynamic FSS limit the time-dependent susceptibility $\chi(t-s) = L^D
\langle M(t)M(s)\rangle$ [$M(t)$ is the instantaneous magnetization at
time $t$ and the average is over all possible dynamic evolutions of
the system in equilibrium] is expected to scale as
\begin{equation}
  \chi(h,L,t) = \chi(h,L,0)\,{\cal X}(\Phi,\Theta)\,.
\label{chi-dynamic}
\end{equation}
The time scale $\tau(L)$ that appears in the definition of $\Theta$
crucially depends on the BC and geometry of the system.  As an
example, we consider an Ising system along the first-order
low-temperature transition line with a small magnetic field $h$.  We
consider PBC and cubic systems of size $L^D$. As we discussed before,
FSS occurs in a small interval of values of $h$, of the order of
$L^{-D}$, and the relevant scaling variable is $\Phi_b$ defined in
Eq.~(\ref{Phib-classical}). At fixed $\Phi_b$ the free-energy
difference between the two magnetized phases is constant for $L\to
\infty$, and typically the system moves between the two magnetized
phases, which are visited with a probability that depends on $\Phi_b$
only. In the coexistence region the slowest dynamic mode is associated
with the flip of the average magnetization, which occurs by generating
configurations characterized by the presence of two coexisting phases
separated by two approximately planar interfaces. Thus, the relevant
time scale is expected to be
\begin{equation}
    \tau(L) \approx L^\alpha e^{2\sigma L^{D-1}}\,,
\end{equation}
where $\sigma$ is the rescaled planar interface tension.  We expect
$\tau(L)$ to depend exponentially on $L$ whenever interfaces are
exponentially suppressed and indeed a similar result holds for OBC
\cite{Fontana-19}. On the other hand, if domain walls are stabilized
by appropriate BC, we expect $\tau(L)$ to scale as a power of $L$,
i.e., $\tau(L) \sim L^\kappa$. This issue was investigated in
two-dimensional Ising and Potts models \cite{Fontana-19,PV-15} using
mixed boundary conditions. In the first direction PBC were used, while
in the second direction different conditions were used on the sides. In
the Ising case, OFBC were used, while in the Potts case, OBC were used
on one side and fixed magnetized BC on the other one.  The exponent
$z$ was estimated, obtaining $\kappa = 2.77(4)$ in the Ising case
\cite{Fontana-19} (cubic systems of size $L^2$) and $\kappa = 3.0(1)$ in
the Potts case \cite{PV-15} (anisotropic strip-like systems of size
$L\times L_\parallel$, with $L_\parallel \gg L$).

The dynamic behavior of systems with O$(N)$ symmetry was discussed in
Ref.~\cite{PV-16,SW-18}.  In this case, because of the continuous
symmetry, no interfaces are present and $\tau(L)$ always scales as
$L^\kappa$. The power $\kappa$, however, depends on the BC and
geometry of the system. If one considers BC that fix the direction of
the magnetization, there are no long-range modes and one obtains
$\kappa = 2$, as appropriate for a dissipative dynamics in an
uncorrelated system.  If instead, the BC do not break the O$(N)$
invariance, the slowest mode is represented by the dynamics of the
average magnetization vector ${\bm M}(t)$, which makes a random walk
on a sphere of radius $m_0(T)$. For cubic geometries, this implies
\cite{PV-16} $\kappa = D$. For cylinder-like geometries $L^{D-1}\times
L_\parallel$ with $L_\parallel \gg L$, we should take into account the
longitudinal correlation length $\xi_\parallel$, introduced in
Sec.~\ref{sec5.2}, which scales as $L^2$ in the FSS limit.  In this
case, one can argue\cite{PV-16} that $\tau(L) \sim L^{D-1}
\xi_\parallel \sim L^{D+1}$.

\subsection{Out-of-equilibrium dynamics: classical quench}

The results of the previous section can be used to predict the
behavior of global observables under a soft classical quench.  In
particular, the two-level reduction we discussed in the quantum case
also holds for classical systems in the absence of domain walls.

We consider an Ising system with $T < T_c$ in a cubic system $L^D$
with PBC, and a magnetized configuration with all spin equal to
$-1$. Then, we consider the evolution of the system in the presence of
a small positive magnetic field $h$. We can then define the average
time-dependent magnetization $M(t)$, where the average is taken over
the different realizations of the dynamics. In the FSS limit
(keeping $\Phi_b=m_0hV/T$ and $\Theta=t/\tau(L)$ fixed), this quantity
satisfies the FSS relation \cite{PV-17-dyn}
\begin{equation}
   M(t,h,L) = m_0 \,{\cal M}(\Phi_b,\Theta) \, .
\end{equation}
The $\Theta$ dependence of this scaling function can be predicted by
using a simpler coarse-grained dynamics, that represents the dynamic
generalization of the two-level truncation we discussed in the quantum
case.  Since on time scales of the order of $\tau(L)$ the reversal of
the sign of the magnetization $M(t)$ is essentially instantaneous, one
can assume that $M(t)$ takes only two values, $\pm m_0$. Second, as
the dynamics restricted within each free-energy minimum is rapidly
mixing, one can assume that the coarse-grained dynamics is Markovian.
Under these conditions, the dynamics is completely parameterized by the
rates $I_+$ and $I_-$ defined by
\begin{eqnarray}
&& P[M(t)=-m_0 \to M(t+dt) = + m_0] = I_+ dt, \nonumber \\
&& P[M(t)=+m_0 \to M(t+dt) = - m_0] = I_- dt, \qquad
\end{eqnarray}
which satisfy ${I_-/I_+} = e^{-2 \Phi_b}$,
to ensure the correct equilibrium distribution. Then, one 
obtains \cite{PV-17}
\begin{equation}
{\cal M}(\Phi_b,\Theta)  = \tanh \Phi_b - (1 - \tanh \Phi_b) e^{-\Theta/T_s},
\end{equation}
where the time scale $T_s$ is a function of $\Phi_b$ only.  These
results are confirmed by the numerical computations reported in
Ref.~\refcite{PV-17-dyn}. They also hold for the thermal transition in
Potts models, apart from the necessary changes due to presence of $q$
equivalent phases on the low-temperature side of the transition
\cite{PV-17-dyn}.

\subsection{Out-of-equilibrium dynamics: Kibble-Zurek scaling}

We finally wish to discuss the Kibble-Zurek
dynamics~\cite{Kibble-80,Zurek-96} in the classical setting, when the
model parameters are slowly varied across the FOCT. For simplicity,
let us consider again an Ising system along the low-temperature
first-order transition line in the presence of a small magnetic field
$h$. We consider the classical analogue of the dynamics discussed in
Sec.~\ref{dynkzfo}.  At $t_i < 0$ we consider an equilibrated
configuration at magnetic field $h_i = t_i/t_s$, Then, we evolve the
system using a diffusive dynamics (for instance, the heat-bath
algorithm), varying $h$ as $h(t) = t/t_s$ and follow the dynamics up
to a time $t_f > 0$. As in the quantum case, we define a FSS regime in
terms of $\Theta = t/\tau(L)$, the same dynamic variable we considered
above, and of a time-dependent quantity $\Phi_{\rm KZ}(t)$, which is
the time-dependent generalization of the FSS variable $\Phi$ defined
in Sec.~\ref{sec5.1}.  We define
\begin{equation}
\Phi_{\rm KZ,\rm cubic} = {m_0\over T} L^D {t\over t_s}\,, \qquad
\Phi_{\rm KZ,\rm cylinder} = {m_0\over T} L^{D-1} 
     \xi_\parallel {t\over t_s}\,, \qquad
\end{equation}
for cubic systems of size $L^D$ and cylinder-shaped systems of 
size $L^{D-1}\times L_\parallel$, respectively. In analogy with what we 
have presented in the quantum case, it is also useful to 
define a scaling variable that is independent of $t$, such as
\begin{equation}
  \Upsilon = {\Theta\over \Phi_{\rm KZ}} =
           {T\over m_0} {t_s\over \tau(L) L^D}\,,
\end{equation}
where the last equality holds for cubic $L^D$ systems.  The FSS limit
is obtained by considering $t,t_s,L\to\infty$, keeping the previous
scaling variables fixed. For simplicity, we keep the initial value
$h_i$ fixed with increasing $L$.  Note that in the limit we have $t_s
\sim L^D \tau(L)$, i.e., $t_s$ is much larger than the typical time
scale of the system, so that the dynamics is effectively able to
equilibrate the system at each time.  In the FSS limit, the
magnetization scales as
\begin{equation}
  M(t,h,L) \approx m_0 \, {\cal M}(\Upsilon,\Phi_{\rm KZ})\,.
\end{equation}
The behavior at thermal transitions is analogous. In this case one
varies the temperature as $\beta(t) = \beta_c (1 + t/t_s)$, so that
$\Phi_{\rm KZ} = (\beta(t) - \beta_c) L^D = \beta_c t L^D/t_s$.  These
scaling behaviors have been verified in magnetic and thermal
transitions \cite{PV-15,PV-16,PV-17}.

\begin{figure}[t]
\begin{center}
\begin{tabular}{cc}
    \includegraphics[width=0.42\columnwidth,angle=0]{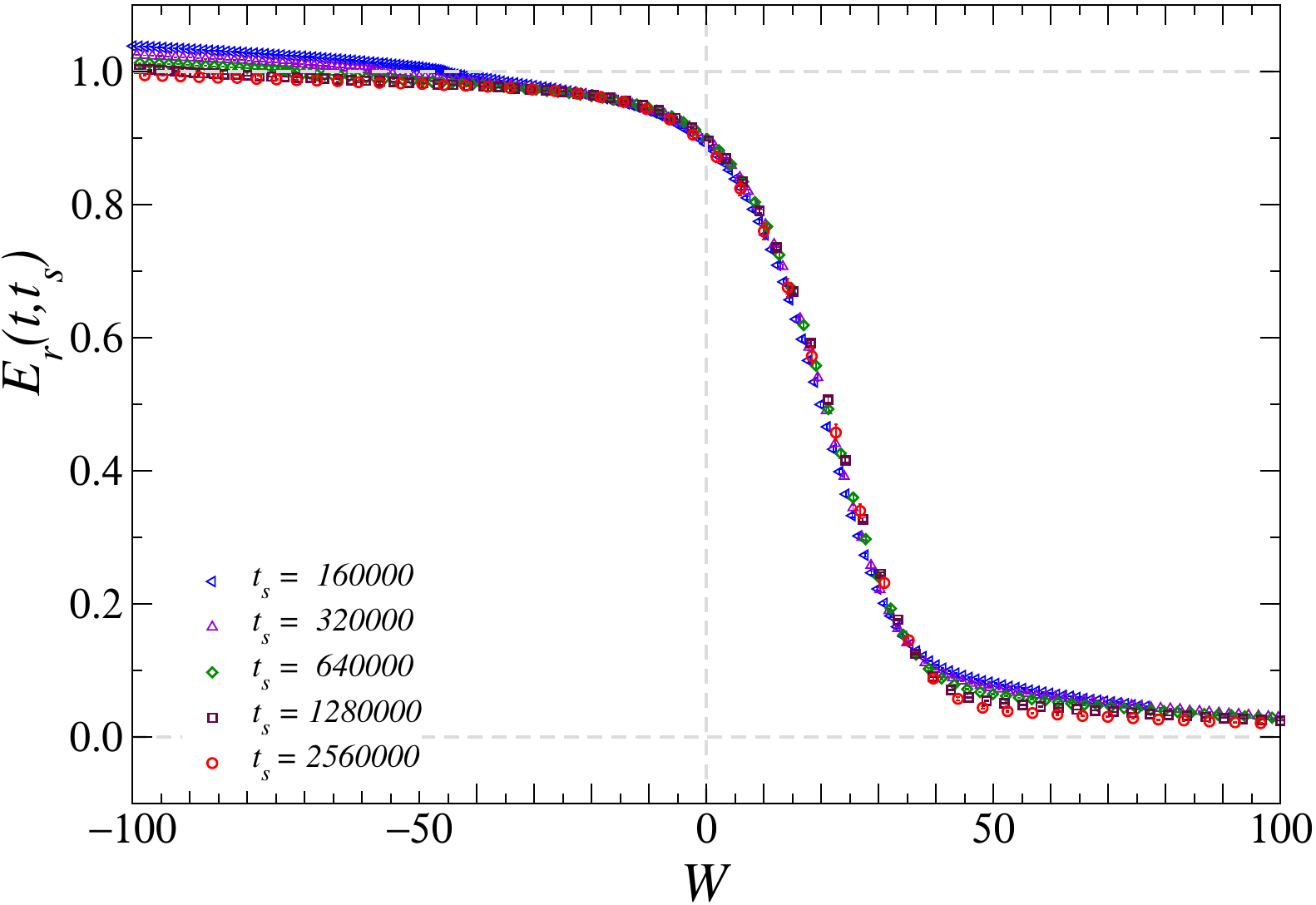} &
    \includegraphics[width=0.48\columnwidth,angle=0]{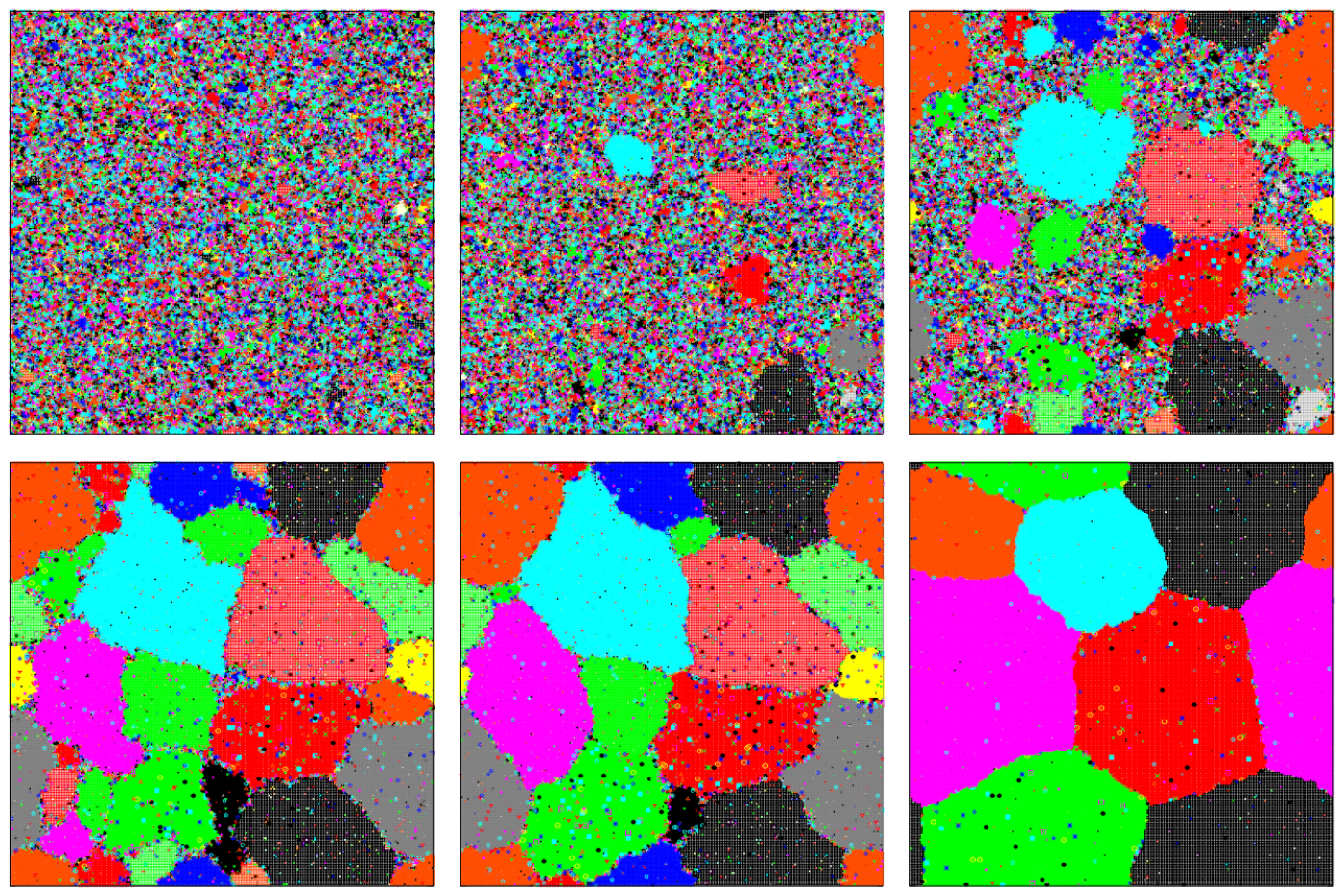}
\end{tabular}
\caption{Results for the two-dimensional Potts model,
  see Eq.~(\ref{cPotts}), with $q=20$ and PBC.  Left: The renormalized energy
  density $E_r$ versus $W$, see Eq.~(\ref{Estar-singKZ}), for several 
  large values of $t_s$.  Right: Snapshots of a
  system of size $L=512$ for $W= -20,\,0,\,20,\,40,\,100,\,1500$
  (from left to right, top to bottom), for $t_s=640000$.  We use
  different colors for each Potts state. Stable droplets
  start appearing for $W\approx 0$, see Ref.~\refcite{PV-17} for
  more details.}
\label{fig:fosym}
  \end{center}
\end{figure}

The Kibble-Zurek dynamics also allows one to investigate the dynamics
of some coarsening phenomena, by considering time scales that satisfy
$t_s \ll \tau(L)$, effectively probing an out-of-equilibrium
infinite-volume regime \cite{PV-17,PPV-18}. Let us briefly review the
results of Ref.~\refcite{PV-17}, for the two-dimensional Potts model
(\ref{cPotts}) with $q=10,20$, in a finite volume with PBC. They considered
a Kibble-Zurek dynamics. The dynamics starts from an
equilibrated high-temperature disordered configuration which evolves
under a purely relaxational heat-bath dynamics with time-dependent
temperature $\beta(t) = \beta_c (1 + t/t_s)$.  They monitored the
instantaneous energy at time $t$, computing its average (with respect
to the thermal history and starting configuration) $E(t,t_s,L)$.  It was
first noted that, for fixed large $t, t_s$, the energy has a
well-defined infinite-volume limit $E_\infty(t,t_s)$. If $L$ is large,
the most important dynamic phenomenon is the droplet formation.  Since
the time needed to create a droplet increases exponentially with its
size $R$ \cite{TB-12}, one can define a length scale $R(t) \sim \ln t$
that gives the typical size of droplets at time $t$. By analogy with
the FSS scaling regime, one may define the scaling
variable~\cite{PV-17}
\begin{equation}
   \Phi_{\rm dr} = \Delta \beta R^D = {t\over t_s} \ln^D t\,,
\end{equation}
and consider the behavior of the system as $t,t_s\to \infty$ at fixed
$\Phi_{\rm dr}$. A numerical analysis for the two-dimensional Potts
model, see the left panel of Fig.~\ref{fig:fosym}, shows that the
energy density obeys the large-$t_s$ asymptotic scaling
behavior~\cite{PV-17}
\begin{equation}
E_r(t,t_s) \equiv  
{E_\infty(t,t_s) - E_-\over E_+ - E_-} \approx  f(W)\,,
\qquad W = (\Phi_{\rm dr}
  - \Phi_{\rm dr}^*) t^\theta\,,
\label{Estar-singKZ}
\end{equation}
where $E_+$ and $E_-$ are the values of the transition-point energies
associated with the the disordered and ordered phases (so that $0\le
E_r \le 1$ in the coexistence region), $\Phi_{\rm dr}^*$ is an
appropriate critical value, and $\theta$ is an exponent characterizing
the dynamics, which turns to be $\theta=1/3$ within the accuracy of
the computations.  Fig.~\ref{fig:fosym} also shows snapshots of 
a configuration for different values of $t$;
it shows droplets of increasing size as the scaling
variable $W$ increases.  The scaling behavior can be interpreted as a
dynamic transition whose position depends on $t_s$.  Indeed, for
each $t_s$ one can compute a time $t_{\rm dr}$ such that $(t_{\rm dr}
\ln^d t_{\rm dr})/t_s = \Phi_{\rm dr}^*$ (this is roughly the time
needed to create a stable large enough droplet), and a corresponding
dynamical (pseudo)-transition temperature $\beta_d = \beta_c ( 1 +
t_{\rm dr}/t_s)$.  The singular behavior (\ref{Estar-singKZ})
resembles the one that occurs at the mean-field spinodal
point~\cite{Binder-87}, but with $\beta_d$ satisfying $\beta_d
-\beta_c \sim (\ln t_s)^{-2} \to 0$. We mention that 
coarsening phenomena in Potts models with OBC were also investigated in
Ref.~\refcite{PPV-18}.  Similar results were obtained
under different dynamic conditions in
Refs.~\refcite{MM-00,LFGC-09,NIW-11,ICA-14,LZ-17,CCP-21,CCEMP-22}.

\end{document}